\documentstyle[epsf,epsfig,latexsym]{mn}

\newcommand{\mnras}[1]{MNRAS}
\newcommand{\apj}[1]{ApJ}
\newcommand{\apjs}[1]{ApJS}
\newcommand{\apjl}[1]{ApJL}
\newcommand{\nat}[1]{Nature}
\newcommand{\aap}[1]{A\&A}
\newcommand{\araa}[1]{ARA\&A}
\newcommand{\aaps}[1]{A\&ASS}
\newcommand{\aj}[1]{AJ}
\newcommand{\apss}[1]{Ap\&SS}

\newcommand{\muy} {~arcsec~yr$^{-1}$}

\newcommand{\kms} {km\,s$^{-1}$}
\newcommand{\ud} {_{\rm disc}}
\newcommand{\uh} {_{\rm halo}}

\author[]{L.V.E.~Koopmans \& R.D. Blandford\\ California Institute of
          Technology, mailcode 130--33, Pasadena CA 91125, USA} 

\date{Accepted; Received} 

\pubyear{2001} 

\title{The Kinematics of High Proper Motion Halo White Dwarfs}

\begin{document}
\maketitle

\begin{abstract}
We analyse the kinematics of the entire spectroscopic sample of 99
recently discovered high proper-motion white dwarfs by Oppenheimer et
al. using a maximum-likelihood analysis, and discuss the claim that
the high-velocity white dwarfs are members of a halo population with a
local density at least ten times greater than traditionally
assumed. We argue that the observations, as reported, are consistent
with the presence of an almost undetected thin disc plus a thick disc,
with densities as conventionally assumed. In addition, there is a
kinematically distinct, flattened, halo population at the more than
99\% confidence level. Surprisingly, the thick disc and halo
populations are indistinguishable in terms of luminosity, color and
apparent age~(1--10~Gyr). Adopting a bimodal, Schwarzschild model for
the local velocity ellipsoid, with the ratios $\sigma_{\rm
U}$:$\sigma_{\rm V}$:$\sigma_{\rm W}$=1:2/3:1/2, we infer radial
velocity dispersions of $\sigma_{\rm U}$=62$^{+8}_{-10}$~\kms\ and
150$^{+80}_{-40}$~\kms\ (90\% C.L.) for the local thick disc and halo
populations, respectively. The thick disc result agrees with the
empirical relation between asymmetric drift and radial velocity
dispersion, inferred from local stellar populations. The local
thick-disc plus halo density of white dwarfs is $n^{\rm td+h}_{\rm
0,WD}$=$(1.9\pm0.5)\times$$10^{-3}$~pc$^{-3}$ (90\% C.L.), of which
$n^{\rm h}_{\rm 0,WD}$=1.1$^{+2.1}_{-0.7}$$\times$$10^{-4}$~pc$^{-3}$
(90\% C.L.)  belongs to the halo, a density about five times higher
than previously thought.  Adopting a mean white-dwarf mass of
0.6~M$_\odot$, the latter amounts to
0.8$^{+1.6}_{-0.5}$$\times$10$^{-2}$ (90\% C.L.) of the nominal local
halo density. Assuming a simple spherical logarithmic potential for
the Galaxy, we infer from our most-likely model an oblate halo
white-dwarf density profile with $n(r)\propto r^{-\alpha}$ and
$\alpha\approx3.0$. The halo white dwarfs contributes $\sim2.6\times
10^{9}$~M$_\odot$, i.e. a mass fraction of $\sim$0.004, to the total
mass inside 50~kpc ($\Omega_{\rm WD}\sim 10^{-4}$). The halo white
dwarf population has a microlensing optical depth towards the LMC of
$\tau^{\rm h}_{\rm WD}\approx1.3\times 10^{-9}$. The thick-disc white
dwarf population gives $\tau^{\rm td}_{\rm WD} \approx4\times
10^{-9}$. The integrated Galactic optical depth from both populations
is 1--2 orders of magnitude below the inferred microlensing optical
depth toward the LMC. If a similar white-dwarf population is present
around the LMC, then self-lensing can not be excluced as explanation
of the MACHO observations. We propose a mechanism that could
preferentially eject disc white dwarfs into the halo with the required
speeds of $\sim$200~\kms, through the orbital instability of evolving
triple star systems.  Prospects for measuring the density and velocity
distribution of the halo population more accurately using the {\sl
Hubble Space Telescope} Advanced Camera for Surveys (ACS) appear to be
good.
\end{abstract} 

\begin{keywords}
stars: white dwarfs -- galaxy: halo -- stellar content -- structure -- 
gravitational lensing -- dark matter
\end{keywords}

\begin{figure*}
\center
\resizebox{0.66\hsize}{!}{\includegraphics{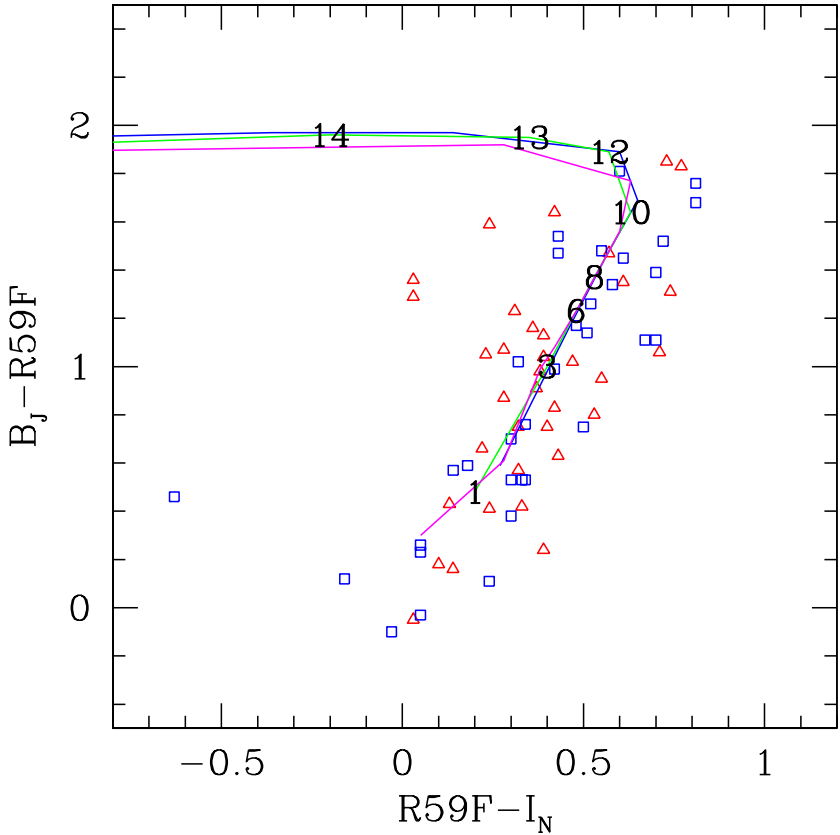}}
\parbox[b]{0.3\hsize}{
\caption{Color-color diagram of the sample of WDs. Shown are the
high-velocity ($>$94~\kms; squares) and low-velocity WDs ($<$94~\kms;
triangles). The velocity cuts in $U$ and $V$ (see Sect.2) are taken as
in Oppenheimer et al.~(2001), for reasons of consistency, but assume
$v_r=0$~\kms. The cooling curves are those from Chabrier et
al.~(2000). The upper to lower curves represent 0.5, 0.6 and
0.8--M$_\odot$ WDs with pure hydrogen atmospheres. Ages in Gyr are
indicated along the 0.5-M$_\odot$ curve. }\vspace{1.2cm}}
\end{figure*}

\section{Introduction}

As the most common stellar remnants, white dwarfs (WD) provide an
invaluable tracer of the early evolution of the Galaxy.  Their current
density and distribution reflects the disposition of their progenitor
main sequence stars and their colors indicate their ages, which are
consistent with many of them having been formed when the Galaxy was
quite young (e.g. Wood 1992).

Until recently, it has been assumed that halo WDs contribute a
negligible fraction of the total mass of the Galaxy (e.g. Gould, Flynn
\& Bahcall 1998). This view is supported by the constraint that the
formation of $\sim$0.6--M$_\odot$ WDs is accompanied by the release of
several solar masses of gas, heavily enriched in CNO elements
(e.g. Charlot \& Silk 1995; Gibson \& Mould 1997; Canal, Isern \&
Ruiz-Lapuente 1997). Yet local stars and interstellar gas, which only
comprise $\sim10$\% of the total Galaxy mass, contain only 2--3
percent of these elements by mass. Based on these metalicity
arguments, most recently Fields, Freese \& Graff (2000) have argued
that $\Omega_{\rm WD}$$<$$3\times10^{-4}$ from C and N element
abundances, adopting H$_0$=70\,km\,s$^{-1}$\,Mpc$^{-1}$.  WDs can
therefore not comprise more than 1--10\% percent of the dark matter in
the Galactic halo within 50~kpc, if $\Omega_{\rm halo}(<50{\rm
~kpc})\sim0.03$ (e.g. Bahcall et al. 2000).  However, this argument
can be circumvented if significant metal outflow from the Galaxy, due
to supernovae, accompanies the formation of these WDs, thereby
removing most of the produced metals from the Galaxy (e.g. Fields,
Mathews \& Schramm 1997).
 
These considerations are important because the MACHO collaboration
(e.g. Alcock et al. 2000) report a frequency of Large Magellanic Cloud
(LMC) microlensing events that is suggestive of 8--50\% (95\% C.L) of
the halo mass in $\sim0.5$--M$_\odot$ compact objects inside a 50~kpc
Galacto-centric radius.  Faint stars and brown dwarfs seem to comprise
only a few percent of the Galaxy mass (e.g. Bahcall et al. 1994; Graff
\& Freese 1996a, 1996b; Mera, Chabrier \& Schaeffer 1996; Flynn, Gould
\& Bahcall 1996; Freese, Fields \& Graff 1998) and are therefore
unlikely to be responsible for these microlensing events. The only
remaining candidates with similar masses, known to us, are WDs.

Alternative explanations for these microlensing events, based on
stellar self-lensing in the LMC have been put forward (Wu 1994; Sahu
1994; Salati et al. 1999; Aubourg et al. 1999; Evans \& Kerins 2000).
A comparable microlensing survey done by the EROS collaboration only
claims an upper limit of 40\% (95\% C.L.) of the halo mass in compact
objects with mass less than 1~M$_\odot$ (Lasserre et al. 2000; see
Afonso et al. 1999 for SMC results).  Because the EROS survey targets
the outer regions of the LMC, in contrast to the MACHO survey, this
could indeed suggest some self-lensing contribution.

There have also been claims of a number of high proper-motion objects
in the HDF, which were suggested to be old WDs in the Galactic halo at
a distance of $\sim$2~kpc (Ibata et al. 1999), with blue colors that
were consistent with revised white-dwarf atmosphere models (Hansen
1999).  More locally, Ibata et al. (2000) found two high proper-motion
WDs.  If the HDF results had been confirmed, they alone would have
been sufficient to explain the observed microlensing optical depth
towards the LMC.  However, recent third-epoch observations have shown
that these ``objects'' were misidentifications and {\sl not} moving
objects (Richer 2001).

The topic has become interesting again, though, with the recent
discovery by Oppenheimer et al (2001) of 38 nearby, old WDs, with
de-projected horizontal velocities in excess of 94~\kms\ (see Sect.2).
They concluded that at least two percent of the halo is composed of
these cool WDs. This conclusion has been challenged, however, by Reid,
Sahu \& Hawley (2001), who claim that these WDs could be the
high-velocity tail of the thick disk. In addition, Hansen (2001) has
argued that these WDs follow the same color-magnitude relation as
those of the disc, which might be unexpected if they are a halo
population.

In this paper, we take the observations of Oppenheimer et al. (2001)
at face value and examine their conclusions in more detail. We assume
that the high proper motion objects have correctly been identified as
WDs and that their inferred distances are statistically
unbiased. However, it is important that additional observations be
performed to validate this.  We also discuss how to set more precise
bounds on the halo WD density.

\section{The White--Dwarf Sample}

The high proper motion WDs were selected from digitized photographic
plates from the SuperCOSMOS Sky Survey\footnote{ see webpage
http://www-wfau.roe.ac.uk/sss/} (Hambly et al. 2001), covering an
effectively searched area of 4165 square degree around the South
Galactic Pole (SGP), upto $b$\,$\approx$$-$45$^\circ$.  Objects with
proper motions between $\mu$=0.33 and 10.0~\muy\ were selected, that
were brighter than $R59F$=19.8 mag and seen at three epochs in each
field. From these objects a sample of 126 potential halo WDs was drawn
based on their reduced proper motion ($H_{\rm
R}$=$R59F$+5\,$\log\mu$+5) and their color $B_{\rm J}-R$.  After
spectroscopic observations of 69 of the 92 WDs without spectra (34 had
published spectra), a sample of 99 WDs remained. For a full discusion
of the sample, its selection criteria and completeness, we refer to
Oppenheimer et al. (2001).

Distances to each of these WDs were ascribed using the photometric
parallax relation derived by Bergeron et al. (1997). This was checked
in two cases using measured distances and the relative uncertainty was
estimated at 20~percent.  Using these distances, Oppenheimer et
al. (2001) subsequently performed a velocity selection on the WD
sample.  The proper motions were converted into a velocity in Galactic
coordinates relative to the local standard of rest, (LSR), with $U$
directed towards the Galactic center (GC), $V$ along the direction of
the circular velocity and $W$ towards the North Galactic Pole (NGP),
see Fig.2. The absence of a measured radial velocity ($v_r$) was
compensated by assuming that $W$=0 and deprojecting onto the $U-V$
plane. It was then argued that 95~percent\footnote{Actually, the
fraction is $\sim86\%$ for a 2--D Rayleigh distribution.} of the
thick-disc WDs have velocities in the plane within 94~km s$^{-1}$
($\approx2\sigma_{{\rm td}}$) from the asymmetric drift point
$(U,V)$=$(0,-35)\mbox{~~km s$^{-1}$}$ and consequently all such WDs
were eliminated, leaving a sample of 38 high-velocity WDs of which 26
are new discoveries and 14 exhibit hydrogen lines and are believed to
be younger. These WDs have velocities in the Galactic plane that are
enclosed within a circle of radius $2\sigma_{{\rm h}}$ centered on
$(U,V)=(0,-220)$~km s$^{-1}$ corresponding to a non-rotating halo
distribution. The density of WDs in the magnitude-limited sample was
then estimated to be $\sim10^{-4}$~M$_\odot$~pc$^{-3}$, roughly ten
times the expected density (e.g. Gould, Flynn \& Bahcall 1998) and
equivalent to 2~percent of the nominal local halo mass density
(e.g. Gates, Gyuk \& Turner 1995).

This procedure has been criticised by Reid, Sahu \& Hawley (2001) who
de-project the velocity vector by setting the radial velocity ($v_r$)
to zero, thereby reducing the number of high velocity WDs.  This
procedure, however, also places several lower velocity WDs outside the
94~km s$^{-1}$ cut. Reid et al. (2001) also note that the high
velocity WDs are not concentrated around $(U,V)=(0,-220)$~km s$^{-1}$
as might be expected from a halo population and argue that the WDs are
mostly associated with the high velocity tail of the thick disc plus a
few WDs from the traditional halo stellar population. Essentially the
same point has been made by Graff (2001). In addition, Gibson \& Flynn
(2001) identified a number of errors in the original table of
white-dwarf properties and argued that the Oppenheimer et al. (2001)
densities are further overestimated by a factor 3--10, bringing them
more into line with the results from other surveys.

In this paper, we analyse the full sample of 99 WDs, not only those 38
with the highest space velocities, thereby avoiding potential pitfalls
associated with arbitrary cuts in the sample and analysing only
subsamples.

\begin{figure}
\begin{center}
\leavevmode
\hbox{%
\epsfxsize=\hsize
\epsffile{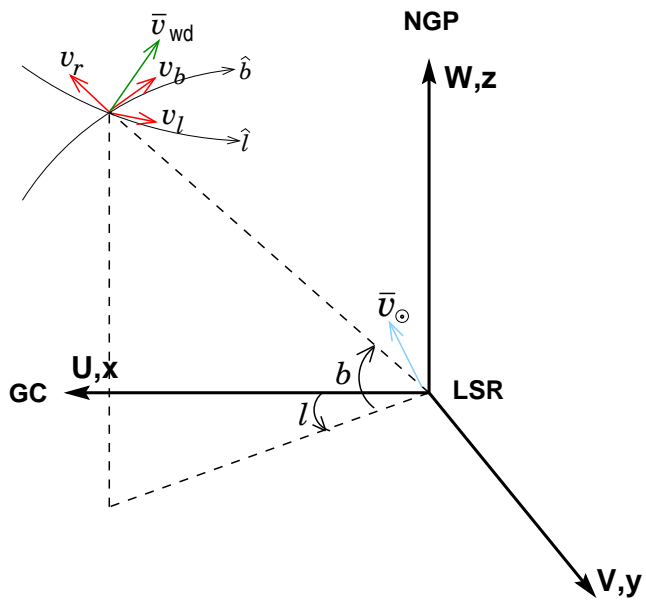}}
\end{center}
\caption{The Galactic coordinate system. The $U$, $V$ and $W$ velocity
components point toward the Galactic center (GC), in the direction of
Galactic rotation (anti-clockwise as seen from the North Galactic
Pole, NGP) and out of the Galactic Plane, respectively. The velocity
components $v_\ell$ and $v_b$ are the projected velocity components on
the sky, in the Galactic coordinates system $(\ell, b)$, of a WD with
space velocity $\vec{v}_{\rm WD}$, and $v_r$ is the radial
velocity. The vector $\vec{v}_\odot$ indicates the Solar motion with
respect to the Local Standard of Rest (LSR).}
\end{figure}

\subsection{White-Dwarf Distances and Ages}

The photometric parallax method to obtain distances hides a potential
problem for the oldest WDs ($\ga$10~Gyr). Although these white dwarfs
continue to cool and fade away, their $B_{\rm J}-R59F$ color remains
nearly constant at $\approx$1.9 for $M_{B_{\rm J}}$$\ga$18 (Fig.1;
e.g. Chabrier et al. 2000). Blindly applying the linear
color-magnitude relation given in Oppenheimer et al. (2001) to the
oldest WDs would therefore lead to a severe overestimate of their
intrinsic luminosity and distance. Even if old WDs are present in the
sample, it might not be surprising that they don't show up in the
color-magnitude relation by Hansen (2001). WDs that move onto the
horizontal branch of the cooling curve have increasingly overestimated
luminositities and consequently move up in the color-magnitude diagram
much closer to the sequence of younger ($\la$10 Gyr) WDs.
Fortunately, only 5\% of the sample of WDs have colors~$B_{\rm
J}-R59F$$\ga$1.8~(Fig.1) and it should therefore not create a major
problem.

However, it is potentially dangerous to create color-magnitude
diagrams, using distances that have been derived by forcing the WDs to
follow the color-magnitude relation of relatively young WDs, and from
those to conclude that the WD are young. This is a circular argument.
A better argument is to plot the age sequence on a two-color diagram
(Fig.1; see also Oppenheimer et al. 2001).  Again, there is no
evidence for WDs older than $\sim10$~Gyr and the high-velocity WDs
appear to be drawn from the same population as the low-velocity WDs
which are presumably relics of ongoing star formation in the
disc. This appears to argue against there being a substantial
white-dwarf halo population (e.g. Hansen 2001), although if there
exists a mechanism that preferentially gives WDs a high space-velocity
during its formation/evolution one might not expect a correlation
between age and velocity. The latter would only be expected if
velocities increased with time due to slower scattering processes. We
return to this in Sect.7.

We should emphasize, however, that the implied ages and distances are
only as good as the assumed underlying white-dwarf color-magnitude
relation, in this case that from Bergeron et al. (1997), and that
independent measurements of their distances are necessary,
preferentially through parallax measurements, even though that will be
difficult given their large proper motions.

\begin{figure*}
\center
\resizebox{0.66\hsize}{!}{\includegraphics{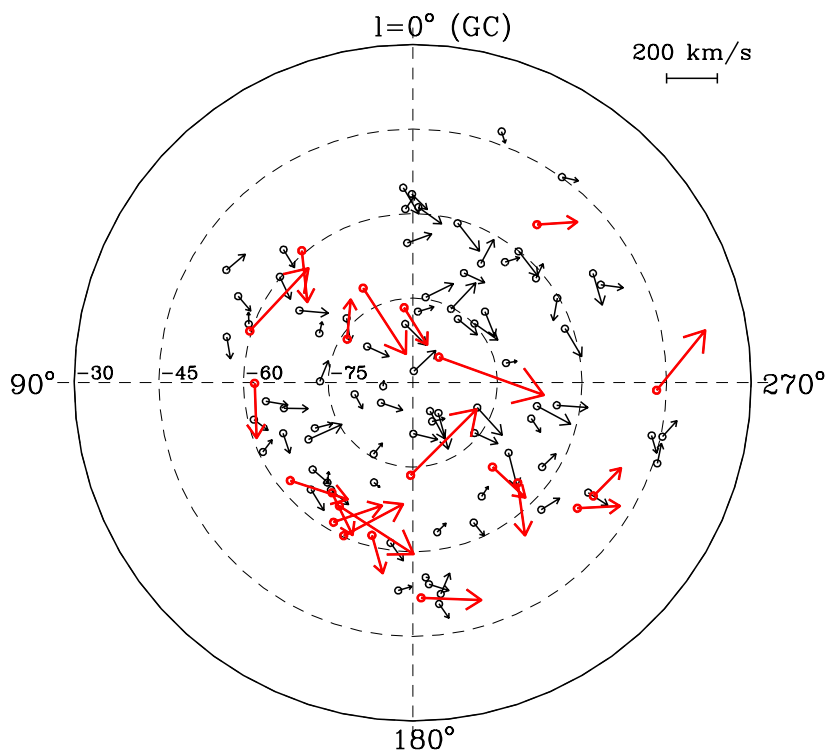}}
\parbox[b]{0.3\hsize}{
\caption{The white-dwarf velocity vectors $(v_\ell,v_b)$, with respect
to an observer near the Sun, projected on the sky in Galactic
coordinates. Galactic latitude increases from $-90^\circ$ to
$-30^\circ$ in steps of $15^\circ$ (dashed circles). Galactic
longitude increase anti-clockwise with $l=0^\circ$ towards to Galactic
center (GC). The arrows indicates the velocity and direction of the
WDs. The asymmetric drift of the higher velocity WDs ($>$150~\kms;
thick arrows) is clearly visible, pointing in the direction opposite
to Galactic rotation ($\ell=90^\circ$).}\vspace{1cm}}
\end{figure*}

%%%%%%%%%%%%%%%%%%%%%%%%%%%%%%%%%%%%%%%%%%%%%%%%%%%%%%%%%%%%%%%%%%%%%%%%%%%%%%%%

\section{Coordinates and Kinematics}

For each WD, we transform its equatorial coordinates
($\alpha$,$\delta$) in to Galactic coordinates ($l$,$b$)\footnote{We
use the transformation in \S 2.1.2 from BM98. Note the typographic
error in $l_{\rm CP}$=122.932$^\circ$ for the Galactic longitude of
the Celestial Pole.}. Using the distances ascribed by Oppenheimer et
al.  2001, we then transform the observed proper motion vector in
Galactic coordinates into a vector $(v_\ell,v_b)$, where $v_\ell$ is
velocity component along the $l$-direction and $v_b$ is the velocity
component along the the $b$-direction (Fig.2). The resulting vector is
orthogonal to the line-of-sight, i.e. the direction of the radial
velocity ($v_r$), which is unknown to us. The transformation from the
observed velocities to Galactic velocities ($U$,$V$,$W$) with respect
to the Local Standard of Rest (LSR), is
\begin{eqnarray}
\label{coordtrans}
U&=&U_\odot + v_r\cos b\cos\ell-v_b\sin b\cos\ell-v_\ell\sin\ell
\nonumber\\ V&=&V_\odot + v_r\cos b\sin\ell-v_b\sin
b\sin\ell+v_\ell\cos\ell \nonumber \\ W&=&W_\odot + v_r\sin b+v_b\cos b
\end{eqnarray}
where $(U_\odot,V_\odot, W_\odot)=(+10,+5.2,+7.2)$~km s$^{-1}$ is the
Solar motion with respect to the LSR (Dehen \& Binney 1998b).  We can
use Eq.(\ref{coordtrans}) to transform the velocity distribution
function $f(U,V,W)$ (\S 4.6) in to $f(v_\ell,v_b,v_r)$, or visa versa,
at the position of any WD with a measured proper-motion vector and
distance.

Fig.3 shows velocity vector ($v_\ell$,$v_b$) for each WD at its
corresponding Galactic coordinates. The velocity vectors point 
preferentially in the direction opposite to Galactic rotation,
which reflects the asymmetric drift (\S 4.2), whereby a stellar population
that is significantly pressure supported, rotates slower than its LSR.
If many of the high proper motion objects had been misidentied on 
the photometric plates, one does not expect a collective movement
of these objects in the same direction.

%%%%%%%%%%%%%%%%%%%%%%%%%%%%%%%%%%%%%%%%%%%%%%%%%%%%%%%%%%%%%%%%%%%%%%%%%%%%%%%%

\section{A Phase--space Density Model}

Before introducing a model for the phase-space density of WDs in the
Solar neighborhood in \S 4.6, we first (i) describe a simple global
potential model that we will employ to derive some of the
characteristic orbital parameters of the sample of WDs, (ii) reprise
epicyclic theory, in order to clarify the discussion and some of the
concepts that we will use in this paper, (iii) derive some
characteristic quantities of the white-dwarfs data set, (iv) examine
the properties of the white-dwarf orbits and (v) discuss some general
properties of the Galactic disc components.

\subsection{A Toy Model of the Galactic Potential}

For definiteness we assume a spherically symmetric potential for the 
Galaxy
\begin{equation}\label{eq:potential}
	\Phi(R,z) = (v_{\rm c}^2/2)\cdot \ln(R^2 + z^2),
\end{equation}
in cylindrical coordinates, where $R$ is the distance from the
rotation axis of the Galaxy and $z$ is the height above the plane.
The simplification of spherical symmetry is not inconsistent with
observations of the Galaxy and external galaxies, even though a wide
range of values for the flattening of the Galactic halo is allowed,
often strongly dependent on the method used to derive it (see Olling
\& Merrifield 2000). This potential, although a clear simplification,
it is to first order also consistent with the relatively flat Galactic
rotation curve (e.g. Dehnen \& Binney 1998a) and allows us to gain
insight in the kinematics and properties of the white-dwarf sample.
The circular velocity $v_{\rm c}$=$\partial \Phi/\partial\ln R$ in
this potential is constant in the plane of the Galaxy ($z$=0).

One can subsequently introduce the effective potential
\begin{equation}\label{eq:potentialeff}
\Phi_{\rm eff}(R,z) = \Phi(R,z) + \frac{L_{\rm z}^2}{2R^2},
\end{equation}
where $L_{\rm z}=R\,v_{\phi}$ is the angular momentum of a WD and
$v_{\phi}$ is its tangential velocity in the plane of the Galaxy. If a
particle is ``at rest'' in the minimum of the effective potential,
i.e. $\partial\Phi_{\rm eff}/\partial R$=0, one readily finds that
$v_{\phi}$=$v_{\rm c}$, hence the WD moves on a circular orbit with a
constant velocity $v_{\rm c}$=$\Omega(R) R$, where $\Omega(R)$ is the
angular velocity.

\subsection{Epicyclic Orbits}

 Most stars and stellar remnants, however, do not move on perfect
circular orbits, but have an additional random velocity and move on
orbits around the minimum in their effective potential. If these
excursions have relatively small amplitudes, the Taylor expansion of
the effective potential around this minimum can be terminated at the
second-order derivative and the stars or WDs will exhibit sinusoidal
(i.e. epicyclic) motions around the circular orbit.  The frequency of
these epicyclic motions are $\kappa^2=\partial^2 \Phi_{\rm eff}/
\partial R^2=(R^{-3}\,{\rm d}(R^4\Omega^2)/{\rm d}R)_{R_{\rm g}}$ in
the tangential and radial directions and $\nu^2=\partial^2 \Phi_{\rm
eff}/\partial z^2$ in the direction out of the plane. The radius of
the guiding centre is $R_{\rm g}$ (see \S 4.2).

\begin{figure*}
\center
\resizebox{0.66\hsize}{!}{\includegraphics{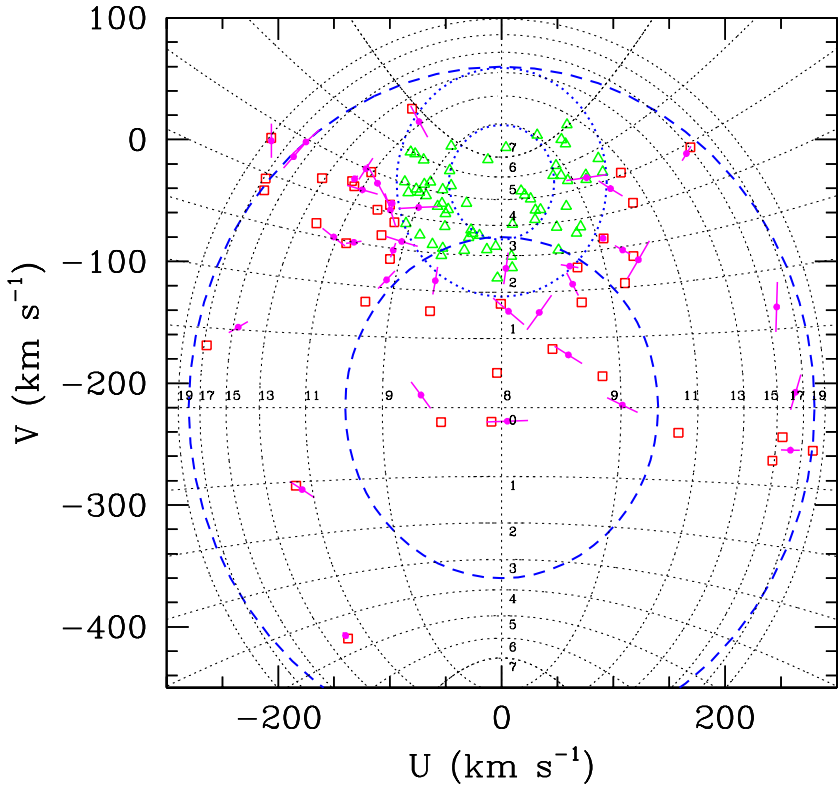}}
\parbox[b]{0.3\hsize}{
\caption{The projection of the 99 white-dwarf velocity vectors on the
U--V velocity plane with respect to the LSR (e.g. Oppenheimer et al.
2001; Reid et al. 2001).  The squares (triangles) indicates the WDs
with velocities in the Galactic plane ($z$=0 and $W$=0) larger
(smaller) than 94~\kms. The thick dotted and dashed circles correspond
to the 1 and 2--$\sigma$ velocity dispersion contours for the disc and
halo populations, respectively.  The projection assumes that
$W$=0. For the population with velocities larger than 94~\kms, also
the $U$ and $V$ velocities are shown under the assumption that
$v_r=0\pm30$~\kms (dots with lines).  The thin dotted ellipsoidal
contours indicate the outermost orbital radii ($R_{\rm max}$, in kpc)
of WDs, whereas the paraboloidal contours indicate the innermost
orbital radii ($R_{\rm min}$, in kpc). It is assumed that R=8~kpc at
the Solar radius and $v_c$=220~\kms.}\vspace{2.0cm} }
\end{figure*}

Because these frequencies depend on local higher-order derivatives of
the potential, we can not use our simple potential model,
Eqn.(\ref{eq:potentialeff}), but have to rely on observational
constraints~(\S 4.2). One can introduce the Oort's constants $A$ and
$B$, whose values have observationally been determined to be
$A$=14.5\,km\,s$^{-1}$\,kpc$^{-1}$ and
$B$=$-$12\,km\,s$^{-1}$\,kpc$^{-1}$ in the Solar
neighborhood\footnote{All subscripts 0 indicate values in the Solar
neighborhood at~$R_0$.} (BT87).  From this is follows that $\kappa_0^2
= -4B(A-B)=-4B\Omega_0$, i.e. $\kappa_0=36$\,km\,s$^{-1}$\,kpc$^{-1}$,
and
\begin{equation}
	{\kappa_0 \over \Omega_0}=2\sqrt{-B \over (A-B)}\approx1.3. 
\end{equation}
The frequency of the epicyclic motion is therefore constant at 1.3
times its orbital frequency, irrespective of the amplitude of the
orbit around the minimum of its effective potential.  In comparison,
the potential given by Eq.(\ref{eq:potential}) gives
${\kappa_0/\Omega_0}=\sqrt{2}\approx1.4$.

\begin{figure*}
\center
\resizebox{0.70\hsize}{!}{\includegraphics{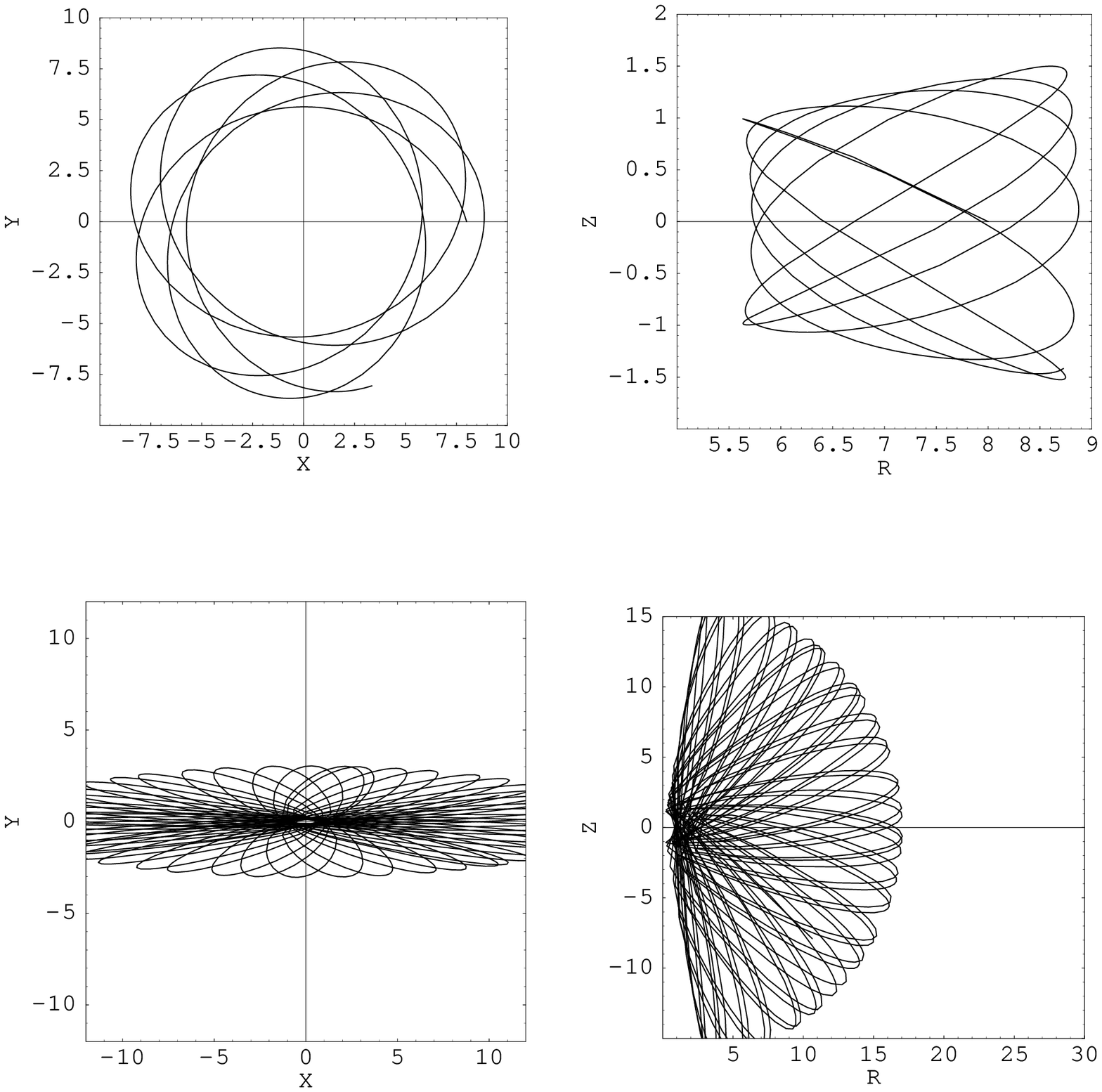}}
\parbox[b]{0.28\hsize}{
\caption{\sloppy Examples of orbits for two WDs drawn from the
sample. (a) High-velocity WD constrained to move in the Galactic plane
with $(U,V,W)= (-264,-169,0)$~\kms. (b) Slow disk WD dwarf with
$(U,V,W)=(49,-21,0)$\,\kms.  (c) The same WD as in (a) except that now
we have assigned the velocities on the assumption that the WD is drawn
at random from the high-velocity component of the bimodal velocity
distribution (\S 4.5). A projection onto the meridional plane is
shown.  (d) The same WD as in (b) with velocities drawn from the slow
component.}\vspace{0.7cm}}
\end{figure*}

Under the epicyclic approximation, stars undergo retrograde,
elliptical, orbits in the plane about their guiding centers, which
move on circular orbits. The ratio of the radial velocity amplitude to
the azimuthal velocity amplitude is $\kappa/2\Omega\approx0.7$, where
$\Omega=27$~km s$^{-1}$~kpc$^{-1}$ is the circular angular frequency
(e.g. BT87) and $\kappa_0$ is given above. The individual orbits are
elongated azimuthally. However, at a given point, retrograde motion by
stars with interior guiding centers that orbit faster than the LSR,
must be combined with prograde motion executed by stars with exterior
guiding centers that orbit more slowly. The net result is that the
ratio of the radial velocity dispersion to the azimuthal velocity
dispersion is predicted to be $\sigma_{\rm U}/\sigma_{\rm
V}$=$2\Omega/\kappa$$\approx$1.5. This prediction is supported by
observations of different stellar populations (e.g. Dehnen \& Binney
1998b; Dehen 1998; Chiba \& Beers 2000).

Superimposed on this horizontal motion is a vertical oscillation with
angular frequency, $\nu=3.6\,\Omega$, which is uncoupled to the
horizontal motion at this level of approximation.  The amplitude of
this oscillation is determined by the past history including
scattering off molecular clouds and spiral arms, and is measured to be
approximately $\sigma_{\rm W}/\sigma_{\rm U}$$\approx$0.5 (e.g. Dehnen
\& Binney 1998b; Dehen 1998; Chiba \& Beers 2000).  The velocity
ellipsoid appears to maintain these dispersion ratios even for
velocities where the epicyclic approximation ought to be quite
inaccurate.

At the next level of approximation, we must include the radial density
variation and this will contribute a mean tangential velocity,
relative to the LSR, at a point, as there are more stars with interior
guiding centers moving backwards than exterior stars moving
forwards. This is the asymmetric drift, $v_{\rm
a}$=$-<$$V$$>$=$v_c-\bar{v}_\phi$, which is the velocity lag of a
population of stars with respect to the LSR. This can be estimated
from the fluid (Jeans') equations by observing that the circular
velocity of a group of stars will be reduced by a quantity
proportional to the pressure gradient per unit density. This should
scale $\propto\sigma_{\rm U}^2$ and it is found (Dehen \& Binney
1998b) that
\begin{equation}\label{eq:asymdrift}
	v_{\rm a}= \frac{\sigma^2_U}{80\pm5\mbox{~\kms}},
\end{equation}
roughly consistent with a calculation using the disc density
variation. Later in the paper (\S 5.1), we use this expression as a
self-consistency check of our results.

In addition, moving groups or non-axisymmetric density variations
(e.g. spiral structure) could be responsible for a second effect, a
rotation of the velocity ellipsoid in the horizontal plane, known as
vertex deviation.  This effect is quite small for old stars
(i.e. $\la$10$^\circ$; Dehen \& Binney 1998b) and we shall therefore
ignore it.

We notice here a key simplification that follows from the realization
that we can observe stars moving with relative azimuthal sky velocity
$V_\perp$ out to a distance $d_{{\rm max}}=V_\perp/\mu_{{\rm min}}$.
By contrast, the amplitude of the radial excursion is $\sim$$2\Omega
V/\kappa^2\sim60 d_{{\rm max}}$. We can therefore ignore curvature of
the stellar orbits and spatial gradients in the stellar density over
the sample volume. A corollary is that we can also ignore a small
rotational correction when relating the proper motion to the space
velocity.

\subsection{Statistical Approach}

There is a standard procedure for finding the mean velocity of a
sample of stars with known proper motions and distances, even when
their distribution on the sky is far from isotropic (BM98). We define
a projection matrix $A_{ij}^n\equiv\delta_{ij}-n_i^nn_j^n$ in Galactic
Cartesian coordinates, where $n^n_i$ is unit vector in the direction
of the n-th star. The velocity that minimises the dispersion about its
projection onto the measured tangential velocity $\vec v_\phi$ is
easily shown to be $[<A>^{-1}]_{ij}<p>_j$. Carrying out this operation
for the full 99 WD sample gives a mean velocity, essentially the
asymmetric drift velocity relative to the LSR, of
$(U,V,W)=(-15,-78,+9)\pm12$~km s$^{-1}$.  This result is fairly robust
in the sense that it is not seriously changed if we remove a few very
high velocities whose reality could be suspect.  The sample has no
significant motion either radially or perpendicular to the plane of
the Galaxy, but clearly shows the asymmetric drift in $V$.  Such an
asymmetric drift, according to Eqn.(\ref{eq:asymdrift}), would naively
imply a velocity dispersion for the complete sample of
$\sigma_U\approx79$\,\kms.  The absence of a velocity drift in the $U$
and $W$ directions also supports the notion that this WD population is
near equilibrium.

The values of the velocity dispersion and asymmetric drift of a
kinematically unbiased sample, however, will be lower than found
above, because a proper-motion limited sample will preferentially bias
toward the high-velocity WDs (see Sect.5).

\subsection{Galactic Orbits}

Because WDs can move distances from the sun that are at least $\sim$2
orders of magnitude larger than their observed distances (\S 4.2), the
majority originate far from the LSR. To quantify this, we have
calculated the inner and outermost radii ($r_{\rm min}$ and $r_{\rm
max}$) of the white-dwarf orbits in the simplified Galactic potential,
Eqn.(\ref{eq:potential}), using the velocities $U$ and $V$, and
explicitly assuming $W$=0 (i.e. the WDs move only in the Galactic
plane).  The assumption $W$=0 leads to a tight scatter of the WDs
around the relation $(U^2+V^2)=(v_\ell^2+v_b^2)$, whereas the
assumption $v_r$=0 leads to results are similar within a few
percent. Statistically, these two different assumptions are therefore
expected to lead to only minor differences.

The results, shown in Fig.~4, indicate that as many as 98, 72, 29 and
14 of the 99 WDs make excursions more than 1, 3, 5 and 7~kpc,
respectively. None of the WDs have both the inner and outermost radii
inside the volume probed by the most distant observed WD (at
$\sim$150\,pc). Hence, even under the restrictive assumption that
these WDs do not move out of the Galactic plane, the majority are not
of local origin, but come from a volume at is at least 3 to 4 orders
of magnitude larger than that probed by the survey.

To illustrate this more graphically, we have plotted the orbits of two
white dwarfs in the sample in Fig.~5, assuming firstly that the WDs
lies in a thin, equatorial disk, i.e. $W=0$, and secondly that they
are drawn from the bi-Maxwellian velocity distribution discussed below
(\S 4.6). One WD has a high velocity and moves on a nearly radial
orbit, the other moves on a nearly circular orbit. It is clear that
the epicyclic approximation is unlikely to provide an accurate
description for the orbits of most high velocity WDs.

\subsection{Disc Models}

In general the disc of our Galaxy is thought to be composed of two
major dynamically distrinct components; the thin disc and a thick
stellar disc, which supposedly comprises several percent of the
stellar disc mass (e.g.  Gilmore \& Reid 1983; Gilmore, Wyse, \&
Kuijken 1989).

Whereas the thin disc has a relatively low velocity dispersion of
$\sigma_W$$\sim$20~\kms\ in the vertical direction and is only several
hundred pc thick, the thick disc has velocity dispersion that is
$\sigma_W$$\sim$45~\kms\ and has an asymmetric drift of $v_{\rm
a}$$\sim$40~\kms (e.g. Gilmore et al. 1989; Chiba \& Beers 2000). If
discs are treated as massive self-gravitating slabs, such that
$\sigma_W^2\sim2\pi G \Sigma z_0$, thick-disc stars typically move
$z_0$$\sim$1~kpc out of the plane for the appropriate Galactic disc
surface densities $\Sigma$, consistent with observations (e.g. Reid \&
Majewski 1993; Robins et al. 1996; Ojha 2001).

Maybe surprisingly, recently analyzed observations suggest (i) that
WDs which kinematically appear to belong to the thin disc, have a
scale height that might be twice that of the traditional thin disc and
(ii) that the WD sky-density towards to North Galactic Pole is an
order of magnitude higher than found in previous surveys (Majewski \&
Siegel 2001).  In comparison, about $\sim$50\% of the high
proper-motion WDs move at least 4~kpc from the LSR (\S 4.4). Hence,
WDs with velocities several times the velocity dispersion of the thick
disc with respect to the LSR could, in principle, move many kpc out of
the Galactic plane as well, if their velocity distribution function is
not extremely flattened in the vertical direction.

WDs observed in the Solar neighborhood are therefore a mixed
population from the thin disc and the thick disc. Because of the
proper-motion cut, we expect only a few WDs from the thin disc in the
sample. If the halo contains a significant population of WDs, it will
also contribute to the local white-dwarf density.  Each of these
populations have their own kinematic properties, although there could
be a continuous transition, for example, from the thick disc to the
halo (e.g. Gilmore et al. 1989).

Whereas, epicycle theory (\S 4.2) indicates that excursions in the
radial and tangential direction are related through conservations of
angular momentum, unfortunately, we have little information about the
velocities perpendicular to the plane.  However, if most WDs are born
in the disc, one expects that scattering processes in the plane of the
galaxy result in comparable radial, azimuthal and vertical
velocities. This is observationally borne out by the fact that the
velocity ellipsoid is isotropic within a factor $\sim$2 (\S 4.2) and
not extremely flattened. Similarly, if WDs occupy the halo as the
result of a merger and/or violent relaxation process, during Galaxy
formation, or if WDs are ejected into the halo from the disc (see \S
7.2) the velocity ellipsoid is also expected to be nearly
isotropic. Theoretically nor observationally is there any evidence or
indication that a velocity distribution can be extremely flattened in
the vertical direction.

If one accepts this conclusion, it is difficult to escape the
consequence that the majority of the observed WDs with large $U$ and
$V$ velocities also have large $W$ velocities and make comparably
large vertical excursions. Even, if we conservatively assume that the
vertical excursions are only half that in the radial and tangential
direction, one finds that $\sim$50\% of these `local' WDs in the
sample are expected to move $\sim2$~kpc out of the plane, twice the
thick disk scale height of $z_0$$\sim$1~kpc, although this conclusion
could be modified if a proper potential model for the stellar disc and
bulge is included (e.g. \S 7.2). In the following subsection we will
examine this in more detail, in the context of a local velocity
distribution function and show more generally how these results can be
translated into a global halo density of WDs.

\subsection{The Schwarzschild Velocity Distribution}

The small sample of WDs and lack of radial velocities does
not allow us to derive the phase-space density of this population,
$F(\bar{x},\bar{v})$, where $\bar{x}$ and $\bar{v}$ are position and
velocity vectors, respectively (e.g. BT87). We must
therefore make some simplifying assumptions: (i) the phase-space
density is locally separable, such that $F(\bar{x},\bar{v})=n(\bar{x})\cdot
f(\bar{v})$, (ii) the velocity distribution function (VDF) with
respect to an observer at the LSR can be described by a normalized
Schwarzschild distribution function (SVDF)
\begin{equation}\label{eq:VDF}
   f_S(U,V,W)=\frac{\exp(-\bar{v}^2/2)}{(2\pi)^{3/2}\sigma_U \sigma_V \sigma_W},
\end{equation}
where~$\bar{v}$=$\left(U/\sigma_U,(V+v_a)/\sigma_V,
W/\sigma_W\right)$ and (iii) $n(\bar{x})$ is constant in the volume probed
by the survey. The adopted coordinates system is shown
in Fig.2. Both the velocity vector ($U$,$V$,$W$), with respect 
to an observer at the
LSR, and the velocity dispersions $\sigma_U$,$\sigma_V$ and
$\sigma_W$\ are defined along the principle axes of the Galactic
coordinate system. In the remainder of this paper, we adopt a Bimodal 
Schwarzschild Velocity Distribution Function (BSVDF) model 
\begin{eqnarray}\label{eq:BSVD}
   f_B(U,V,W) &= &\frac{r\,\exp(-\vec{v}_{\rm td}^2/2)}{(1+r)(2\pi)^{3/2}\sigma_{U,\rm td} 
	 	\sigma_{V,\rm td} \sigma_{W,\rm td}}~ + \nonumber \\
	      &  &\frac{\exp(-\vec{v}_{\rm h}^2/2)}{(1+r)(2\pi)^{3/2}\sigma_{U,\rm h} 
	\sigma_{V,\rm h} \sigma_{W,\rm h}},
\end{eqnarray} 
where the definitions are as in Eqn.(\ref{eq:VDF}) and $r=n_{\rm td}/n_{\rm h}$ is the
number-density ratio of WDs in the thick disc
and halo. The velocity dispersions, asymmetric
drifts and number-density ratio are free parameters. We emphasize that the adopted BSVDF 
model is not a unique choice and the real VDF could be a continuous
superposition of SVDFs or have a form different from Maxwellian. Experiments with more
general forms of the VDF (power laws), however, give essentially similar results.
In addition, the SVDF is locally a solution of the collisionless Boltzmann equation 
and also internally consistent with epicyclic theory (BT87). The same is true for 
a superposition of SVDFs.

\section{Maximum Likelihood Analysis}

To calculate the probability of a set of  model parameters,
$M$=\{$\sigma_{U,\rm d}$,$\sigma_{U,\rm h}$,$v_{a,\rm
d}$,$v_{a,\rm h}$,$r$\}, given the set of constraints $D_i=\{\mu_i,pa_i\}$,
i.e. the proper motion $\mu$ and position angle $pa$ from the $i$-th white
dwarf, we use Bayes' theorem
\begin{equation}
	P_i(M|D_i)=\frac{P(M)\,P(D_i|M)}{P(D_i)}.
\end{equation}
We assume that $P(M)$ is a constant and set it equal to unity,
although this might not the case. In particular, the efficiency of
detecting WDs might be a function of its proper motion and position on
the sky (Oppenheimer, private communication). If the values of
$P(D_i|M)$ are statistically independent, the log-likelihood sample
becomes
\begin{equation}
	{\cal L}(M|D)=\sum_{i=1}^{N_{\small \rm WD}} \ln P(D_i|M),
\end{equation}
where $D=\{D_i\}$ and $N_{\small \rm WD}$ is the number of WDs in the
sample.  The probability $P(D_i|M)$ for each WD in a {\sl fully}
proper-motion limited is given by the function
\begin{equation}\label{eq:PDM}
	P(D_i|M) = \frac{\int_\infty^\infty W(v_l,v_b)
	f(\vec{v}=\vec{p}_i+v_r\hat{r}_i)\, {\rm d}v_r} {\int_{\cal V}
	{\rm d}^2v \int_\infty^\infty W(v_l,v_b)
	f(\vec{v}=\vec{p}_i+v_r\hat{r}_i)\, {\rm d}v_r},
\end{equation}
where $\vec{p}_i$=$(v_\ell,v_b)_i$ is the velocity vector on the sky
in Galactic coordinates (see Eqn.1) and $\hat{r}_i$ is the unit vector
in the radial direction away from the observer (Graff and Gould,
private communications). The weight function $W(v_l,v_b)=(v_l^2 +
v_b^2)^{3/2}$ and $\cal V$ indicates that area in velocity space for
which $(v_l^2 + v_b^2)>(\mu_{\rm min} r_i)^2$, where $r_i$ is the
distance to the WD and $\mu_{\rm min}=0.33$\muy\ is the proper-motion
lower limit of the survey. The weight function is proportional to the
maximum volume ($V_{\rm max}$; see Sect.6) in which the WD could have
been found and therefore biases a proper-motion limited survey towards
high-velocity WDs. In case $\mu_{\rm min}=0$, the likelihood function
reduces to that of a kinematically unbiased survey (e.g. Dehnen 1998).

In practice, only $\sim$90\% of the WDs in the sample are
proper-motion limited, whereas only $\sim$10\% of the observed WDs
would first drop out of the sample because of the survey's magnitude
limit (Fig.7; see Sect.6 for a more detailed discussion). These WDs
are also those with highest observed velocities (Fig.7). If we assume
that these WDs are also proper-motion limited, we would overestimate
the volume (i.e. $W(v_l,v_b)$) in which these WDs could have been
detected and thus bias the results towards lower velocities and lower
halo densities (Gould, private communications), the latter by about 
40--50\%. Of the ten WDs that are magnitude limited, eight have
nearly identical values of $V_{\rm max}$ (and therefore absolute
magnitude), whereas the other two have much smaller
values. Conservatively, we assume that all ten WDs have similar and
constant values of $V_{\rm max}$, which might slightly underestimate
the halo density of WDs. In practice this effect is only $\sim$10\%
and we therefore neglect it in light of all the uncertainties.  It is
important to remember, however, that strictly speaking Eq.(\ref{eq:PDM})
is only valid for completely proper-motion limited surveys. The notion
that this is the case for $\sim$90\% of the WDs allows us to make the
simple adjustment above without explicit knowledge of the white-dwarf
luminosity function, which is roughly speaking assumed to be a delta
function.

For each WD, $i$, we now know the velocity vector $\vec{p}_i$ (see Fig.3). 
We then evaluate $P(D_i|M)$ using Eqn.(\ref{eq:PDM}) and the
transformation between $(v_\ell,v_b,v_r)$ and $(U,V,W)$, as discussed
in Sect.3 and from those we can evaluate ${\cal L}(M|D)$, given a set 
of model parameters $M$.

\subsection{Results from the BSVDF model} 

We maximize the likelihood ${\cal L}(M|D)$, given the BSVDF model
introduced in \S 4.6 and the proper-motion data set from the 99
observed WDs. Before the optimisation, we first remove three of the
WDs that contribute most ($\sim$80\%) to the WD density determined
from the complete sample. A justification for this is given in
Sect.6. The removal changes the results only a few percent compared to
the dataset of 99 WDs. The resulting dataset from 96 WDs is designated
$D_{96}$.  We optimize by varying all five parameters
$M$=\{$\sigma_{U,\rm d}$,$\sigma_{U,\rm h}$,$v_{a,\rm d}$,$v_{a,\rm
h}$,$r$\}. For definiteness we assume that the velocity ellipsoid has
ratios $\sigma_U$:$\sigma_V$:$\sigma_W$=1:2/3:1/2, close to what is
observed for the smooth background of stars in the local neighborhood
(e.g. Dehnen 1998; see \S 4.2). The resulting model parameters are
listed in Table~1. The errors indicate the 90\% confidence level,
determined by re-optimizing, whereby we allow all parameters to vary,
but keep the parameter of interest fixed.

\begin{table}
\caption{The most likely BSVDF model for the full ($D_{96}$). 
The errors indicate the 90\% confidence interval.}
\begin{center}
\begin{tabular}{ll}
Parameter & $D_{96}$   \\
\hline
$\sigma_{U,\rm d}$ (\kms) &  62$^{+8}_{-10}$         \\
$\sigma_{U,\rm h}$ (\kms) & 150$^{+80}_{-40}$        \\
$v_{a,\rm d}$ (\kms)      &  50$^{+10}_{-11}$        \\
$v_{a,\rm h}$ (\kms)      & 176$^{+102}_{-80}$       \\
$r=n\ud/n\uh$             & 16.0$^{+30.0}_{-10.5}$   \\
\hline
\end{tabular}
\end{center}
\end{table}

Fig.6 shows the likelihood contours of thick-disc and halo velocity dispersions 
around the most likely models in Table~1. We maximize the likelihood for 
each set of velocity dispersions. In the case $D_{96}$, 
one can conclude that VDF is bimodal at the more than 99\% C.L. 

\begin{figure}
\begin{center}
\leavevmode
\hbox{%
\epsfxsize=\hsize
\epsffile{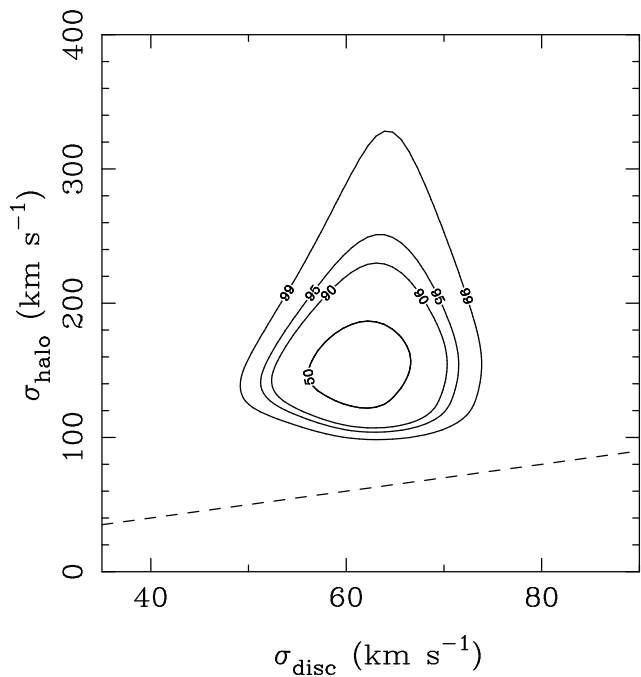}
}
\end{center}
\caption{Likelihood contour plot of $\sigma_{\rm disc}$ versus
$\sigma_{\rm halo}$ for the full data set (96 WDs). The dashed
line indicates $\sigma_{\rm disc}=\sigma_{\rm halo}$.} 
\end{figure}

Due to the position of the WDs close to the South Galactic Pole, only
poor constraints are obtained on the $\sigma_W$. Our particular choice
of the ratio $\sigma_U:\sigma_W$=$1:1/2$ could potentially
underestimate the density of WDs in the halo.  By varying the ratio
$(\sigma_W/\sigma_U)$ and re-optimizing the likelihood, we find
$(\sigma_W/\sigma_U)<1.1$ at the 90\% C.L., with no sensible lower
limit.  This exemplifies the poor constraints on velocities
perpendicular to the Galactic plane. Within this range the parameters
in Table~1 vary by only a few percent.

Note also that the maximum-likelihood estimate of the thick-disc
asymmetric drift, $v_{a,\rm d}$=50~\kms, agrees well with that
estimated from the asymmetric-drift relation in Eqn.(5), which gives
48~\kms. If we make the velocity vector isotropic, for example, this
is no longer the case and the resulting asymmetric drift and radial
velocity dispersion of the thick-disc population
(i.e. $\sigma_U$=51~\kms\ and $v_a$=49~\kms) are inconsistent with the
empirical relation in Eqn.(5). The likelihood of the model also
decreases by a six orders of magnitude. This might be regarded as a
consistency check that the sample of thick-disc WDs and the likelihood
results are in agreement with the kinematics of other local stellar
populations, whereas this is not the case for an isotropic velocity
distribution. The flattening of the velocity ellipsoid in the $V$
direction also naturally explains the apparent scarcity of WDs in the
regions around $(U,V)=(0,-220)$~\kms\ and $V<-220$~\kms, noticed by
Reid et al. (2001).

%%%%%%%%%%%%%%%%%%%%%%%%%%%%%%%%%%%%%%%%%%%%%%%%%%%%%%%%%%%%%%%%%%%%%%%%%%%%%%%%

\begin{figure}
\begin{center}
\leavevmode
\hbox{%
\epsfxsize=\hsize
\epsffile{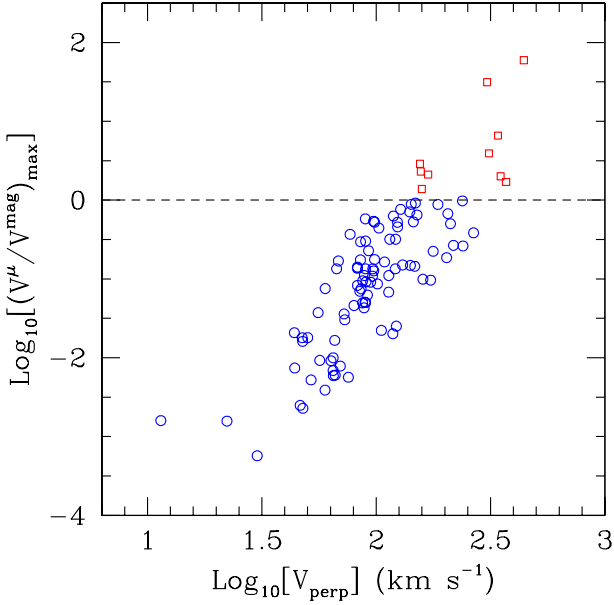}
}
\end{center}
\caption{The ratio $V^{\mu}_{\rm max}/V^{\rm mag}_{\rm max}$ as function
of observed velocity on the sky ($v^2_{\perp}$$\equiv$$v_\ell^2 +
v_b^2$). The ten WDs (squares) with $V^{\mu}_{\rm max}/V^{\rm mag}_{\rm
max}$$>1$ are magnitude-limited in the sense that they would still be
in the proper motion sample if they were at a distance such that their
apparent magnitude equaled the survey limit. The remaining 89 WDs
(circles) are proper-motion limited. Note that most of the
magnitude-limited WDs have high velocities and low luminosities. A
linear fit to sample shows a dependence $\propto$$v_\perp^3$, as
expected for a proper-motion limited sample with no correlation
between white-dwarf luminosity and velocity.}
\end{figure}

\begin{figure}
\begin{center}
\leavevmode
\hbox{%
\epsfxsize=\hsize
\epsffile{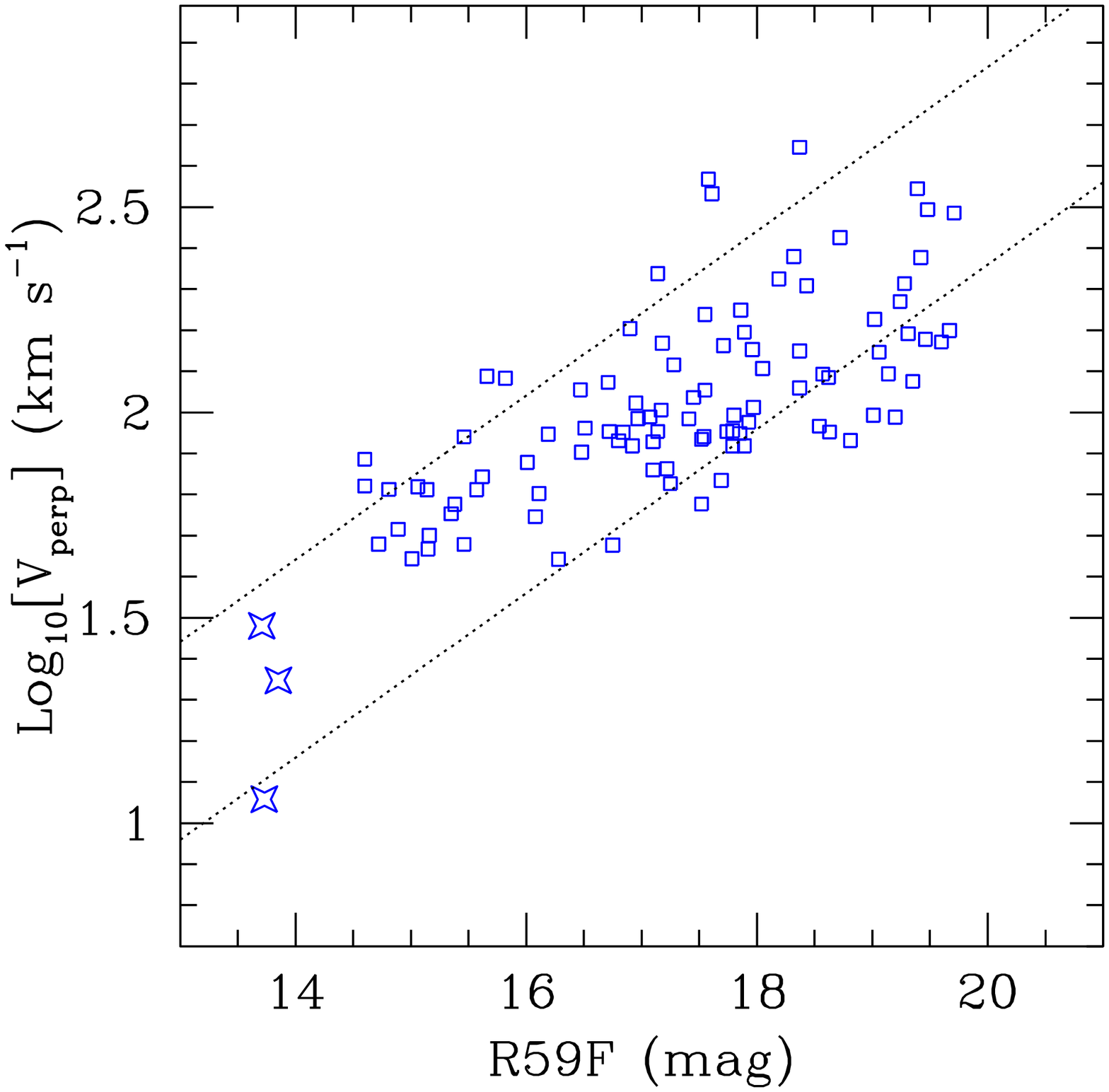}}
\end{center}
\caption{The correlation between velocity on the sky ($v_\perp$) and
magnitude $R59F$. The lower dotted line results from the relations
$V^{\rm m}_{\rm max}=V^{\mu}_{\rm max}$, $\mu_{\rm min}$=0.33~\muy and
a constant absolute WD magnitude $<M_{R59F}>$=14.3 mag, i.e. the
average of the sample of WDs.  Below this line, WDs are either too
faint or slow to be detected. Some WDs are brighter than
$M_{R59F}$=14.3 mag and lie below the line, but in general most stars
follow this lower limit. The upper dotted line indicates the same, but
for $\mu_{\rm min}$=1\muy. The local WD density is dominated by the 3
WDs (stars) that lie isolated from the bulk of the population, and
which presumably belong to the thin-disc (see text).}
\end{figure}

\section{The Local White Dwarf Densities}

We must now turn to estimating the local space density which is needed
to normalize the stellar distribution function. Our approach follows
that of Oppenheimer et al (2001) by taking into account the proper
motion selection, as well as the flux selection. We continue to ignore
errors, which can introduce significant biases, and suppose that a
star with proper motion $\mu$, magnitude $R59F$ and distance $d$,
could have been seen out to a distance $d_{{\rm max}}=\min[d^\mu_{{\rm
max}},d^m_{{\rm max}}]$ where the proper-motion distance limit is
$d^\mu_{{\rm max}}= v_\perp/\mu_{\rm min}$, with $\mu_{\rm
min}$=0.33~\muy, and the magnitude distance limit is given by
$\log(d^m_{\rm max})=1+(19.8-R59F)/5$ and still been included in the
sample.

The maximum volumes in which the WD could have been found are then
$V^{\mu/m}_{\rm max}=(\Omega_{\rm s}/3)\times (d^{\mu/m}_{\rm
max})^3$, where $\Omega_{\rm s}$ is solid angle covered by the survey.
In Fig.7, we plot the ratio $V^{m}_{\rm max}/V^{\mu}_{\rm max}$ versus
observed velocity ($v_\perp$). One observes that 89 out of the sample
of 99 WDs are proper motion-limited; there is a conspicuous absence of
faint WD in the sample relative to a true magnitude-limited
sample. Note that the absence of faint WDs is even more prominent for
the slowly moving stars (Fig.8). There are two possible
explanations. Either the photometry is inaccurate and the selection is
quite incomplete close to the survey limit $R59F=19.8$. Alternatively,
the density of old, low luminosity WDs is so small that there are no
more than a few that are close enough for us to see them move. For
this second explanation to be viable, the logarithmic slope of the
luminosity function $\Phi(L)$ must be greater (and in practice much
greater) than $-5/2$.  In order to distinguish between these two
explanations we have evaluated $<V/V_{{\rm max}}>= <(d/d_{{\rm
max}})^3>$ for the whole sample. We find a value $<V/V_{{\rm
max}}>=0.41$. This is quite close to the expect value of 0.5 that we
suspect that although incompleteness is present, the distribution in
$d-v-L$ parameter space reflects the whole population, where $L$ is
the luminosity of the WD.

If we ignore possible photometric incompleteness, our estimator of the
total space density is
\begin{equation}
      n^{\rm total}_{0,{\rm WD}}={3f_\Omega^{-1}f_S^{-1}
	\over4\pi}\times \sum_{i=1}^{N_{\rm WD}} {d_{{\rm max,i}}^{-3}}
\end{equation}
where $f_\Omega=0.12$ is the fraction of the sky surveyed and
$f_S=0.8$ is the estimated spectroscopic incompleteness of the 99 WD
sample.  The result is $n^{\rm total}_{0,{\rm WD}}=0.0096$~pc$^{-3}$.
The result is quite robust with respect to changes in the limiting
magnitude although 30\% of the density is contributed by a single WD,
and 80\% by three WDs (i.e those with $R59F<14$ mag; Fig.8).  This
estimator refers to stars whose luminosities and transverse velocities
are actually included in the sample although the variance can be quite
large if there are significant contributions from WD near the
boundaries of this distribution. From Fig.1 it is apparent that this
range translates into absolute magnitudes, $16\la M_{R59F}\la$19 and
ages in the range of about $1-10$~Gyr.

The three brightest WDs that dominate the density estimate are most
likely part of the thin-disc population. Because this component was
not included in the VDF model used in the likelihood analysis,
we have to remove these objects from the density estimate to avoid a
severe overestimate of $n^{\rm td+h}_{0,{\rm WD}}$, the normalisation
of the local phase-space density, $F_{B}(U,V,W)=n^{\rm td+h}_{0,\rm
WD}\times f_{B}(U,V,W)$, of the thick-disc and halo WDs.  We therefore
remove all WDs with observed velocities $v_\perp<30$~\kms\ from the
sample (three in total), which should remove all thin-disc WDs.  Their
removal from the sample has a negligible effect on the likelihood
analysis (see \S 5.1).  We subsequently find ${n}^{\rm td+h}_{0,{\rm
WD}}=(1.8\pm0.3)\times10^{-3}$~pc$^{-3}$ (1--$\sigma$), 
which then represents the local
density of thick-disk and halo white dwarfs that have
$v_\perp\ge30$~\kms.

To normalise $F_{B}(U,V,W)$ we calculate the fraction, $f_{>30}$, of
the VDF $f_{B}(U,V,W)$ for which the observed velocity
$v_\perp\ge30$~\kms. The correct normalisation of the local density of
thick-disk and halo WDs then becomes $n^{\rm td+h}_{0,{\rm WD}}=
\tilde{n}^{\rm td+h}_{0,{\rm WD}}/f_{>30}$. Given the BSVDF model
listed in Table~1, for the full data-set ($D_{96}$), we find
$<f_{>30}>=0.946$, hence
$$n^{\rm td+h}_{0,{\rm WD}}=(1.9\pm0.5)\times10^{-3}~{\rm pc^{-3}~~~~(90\% ~C.L.)}.$$
Finally, the {\sl local} halo density of WDs becomes
$$
n^{\rm h}_{0,{\rm WD}}={n^{\rm td+h}_{0,{\rm WD}} \over (1+r)}=
1.1^{+2.1}_{-0.7} \times 10^{-4} ~{\rm pc^{-3}~~(90\% ~C.L.)}
$$
with $r$=16.0$^{+30.0}_{-10.5}$ (Table~1).  This density estimate for
the halo white dwarfs is about half that found by Oppenheimer et
al. (2001).  For completeness, we quote a thick-disc white-dwarf
density of
$$
n^{\rm td}_{0,{\rm WD}}={n^{\rm td+h}_{0,{\rm WD}} r \over (1+r)}=
(1.8\pm0.5)\times 10^{-3} {\rm ~pc^{-3}~~(90\% ~C.L.)},
$$
which corresponds within $1.5\,\sigma$ to the thick-disc number
density $n^{\rm td}_{0,{\rm WD}}=(3.3\pm1.2)\times 10^{-3}$~pc$^{-3}$
that was estimated by Reid et al. (2001) on the basis of only seven
WDs within a sphere of 8~pc from the Sun. Note that our estimate takes
the contribution from WDs outside the 94~\kms\ cut in to account. We
therefore self-consistently recover the local thick-disc density
within acceptable errors, from the sample of high-proper motion WDs
discovered by Oppenheimer et al. (2001).  We therefore do not expect
this sample to be significantly incomplete or biased in a way unknown
to us.

Concluding, the likelihood result that there are two kinematically
distinct populations, at the 99\% C.L, gives us confidence that there
is indeed a contribution of $\sim$6\% of the thick-disk WD density or
$\sim$0.8\% of the local halo density, by a population of WDs that
exhibits kinematic properties that are not dissimilar to that expected
from a nearly pressure-supported halo population.

\begin{figure}
\begin{center}
\leavevmode
\hbox{%
\epsfxsize=\hsize
\epsffile{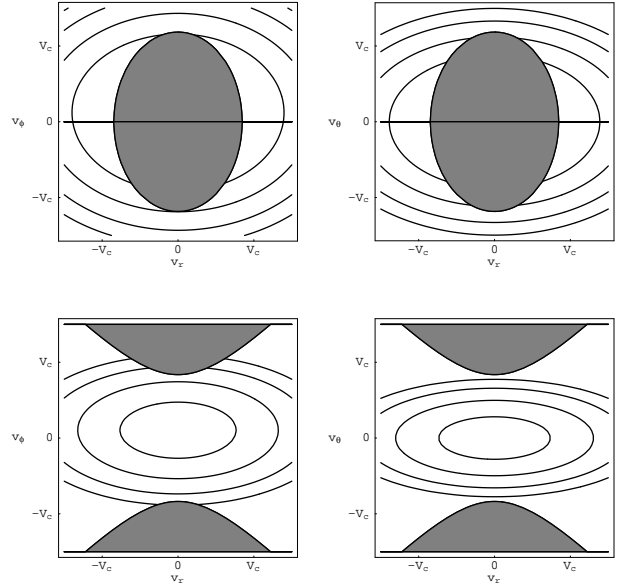}
}
\end{center}
\caption{Three dimensional WD velocity distributions at two locations.
The top row corresponds to radius $r=0.7\,r_c=5.6$~kpc and colatitude
$\theta=60^\circ$ relative to the Galactic center and the bottom row
to $r=1.4\,r_c=11.2$~kpc, $\theta=60^\circ$. The projections on the left
are of $f(v_r,0,v_\phi)$ and those on the right are of
$f(v_r,v_\theta,0)$. The velocity scale $v_c=220$~km s$^{-1}$ is
marked. The four contour levels are at $0.01,0.03,0.1,0.3$ times the
local peak halo WD distribution function, $f_0=1.5\times10^{-11}{\rm
pc}^{-3}({\rm km s}^{-1})^{-3}$. The shaded areas are excluded volumes
in velocity space where, under the dynamical assumptions discussed in
the text, local measurements of the WD distribution cannot inform us
about the halo distribution.}
\end{figure}

\subsection{Local Stellar and Halo Densities} 

To place the results obtained above in perspective, we now turn to a
comparison of the local density of WDs that we derived with other local
densities. 

Summarising, the local mass densities of WDs, derived in this 
paper, were $\rho_{0,{\rm WD}}\approx 0.7\times 10^{-4}$, $1.1\times 10^{-3}$ and
$6\times10^{-3}$~M$_\odot$~pc$^{-3}$, respectively, for the halo,
thick-disc and total mass density (including thin disc), if we convert
number densities to densities using an average WD mass of 0.6~M$_\odot$.
The local
WD density derived from direct observations is $(3.7\pm0.5)\times
10^{-3}$~M$_\odot$~pc$^{-3}$ (Holberg, Oswalt \& Sion 2001). These direct observations
compare well with our total WD density estimate. Because of the very small volume 
probed by the proper-motion limited survey of Oppenheimer et al. (2001) 
for the lowest velocity thin-disc WDs, the error on our estimate is
very large. We found $\sim$3 plausible thin-disc WDs (Sect.6; Fig.8). If we
had found one WD less, with comparable or lower velocities, 
the total density estimate could easily have halfed. The 
low number statistics of thin-disc WDs, which dominate the local WD
density, makes the integrated density estimate of local WDs very uncertain. 
This is much less of a problem for both the thick-disc and halo density
estimates, because in both cases the probed volume is significantly
larger (2--3 orders of magnitude) and consequently also the number
of WDs from both populations. 

The local stellar density is around $0.06-0.1$~M$_\odot$~pc$^{-3}$ (e.g.
Pham 1997; Creze et al. 1998; Holmberg \& Flynn 2000). The mass density
ratio of WDs to stars is therefore about 4--6\%. This fraction 
is 2--3 times less than the $\sim$13\% estimated by Hansen (2001). We find 
$\sim$15\% from a more sophisticated population synthesis model that
self-consistently takes metallicity-dependent yields, gas in- and 
outflow, stellar lifetimes and delayed mixing into account, under the 
same assumption that the stellar IMF is Salpeter from 0.1 to 100~M$_\odot$.
The result is almost independent of any assumption, except for the 
shape and lower mass limit of the IMF. We return to this apparent 
discrepancy in \S 7.2. 

The local halo density in dark matter is estimated to be
$(8\pm3)\times 10^{-3}$~M$_\odot$~pc$^{-3}$ (Gates et al. 1995; where
we take the result that does {\sl not} include the microlensing
constraints).  This result agrees with those determined by other
authors.  Our local halo WD density estimate would therefore
correspond to $\sim$0.8\% of the local halo mass density, slightly
higher than that estimated by Oppenheimer et al. (2001), for the
reasons outlined previously.

The local density in hydrogen-burning halo stars is estimated to be
$(4\pm1)\times 10^{-5}$~M$_\odot$~pc$^{-3}$, whereas the WD density
estimated from this population, based on population synthesis models,
is $1.3\times10^{-5}$~M$_\odot$~pc$^{-3}$ (Gould et al. 1998).  Our
density estimate of WDs is therefore $\sim$5 times larger than that
previously estimated and is comparable in mass to the local stellar
halo density. Such a high density in halo WDs demands a non-standard
explanation, other than that based on simple population synthesis
models and/or a standard IMF. We discuss this in more detail in \S
7.2.

First, however, we examine the contribution of these high-velocity WDs
to the non-local halo mass density, which allows us to estimate their
total mass contribution to the Galaxy, under some simplifying
assumptions.

\section{The Halo White Dwarf Density}

Combining the results of Sections 5 and 6 we now have a normalised
halo distribution function in the form
\begin{equation}
F_h(\vec{x}_0,\vec v)=n_{0,\rm WD}^{\rm h} \times {\exp(-\vec
v^2/2)\over(2\pi)^{3/2}\sigma_U\sigma_V\sigma_W}
\end{equation}
evaluated at the solar radius ($r_c=8$~kpc). What does this tell us
about the total white-dwarf density throughout the halo? The
connection is provided by the collisionless Boltzmann (or Vlasov)
equation which states that the distribution function does not vary
along a dynamical trajectory. Therefore, if we knew the halo potential
$\Phi(\vec r)$ as well as the local velocity distribution function, we
would know the halo distribution function at radius $r$ and colatitude
$\theta$ for all velocities that connect to the solar radius. (We
assume that the distribution functions is symmetric.)

We now make a further simplification and assume that the Galactic
potential, $\Phi(\vec r)$=$\Phi(|\vec r|)$, is spherically symmetric.
Specifically, we adopt Eqn.(\ref{eq:potential}) although our results
are not particularly sensitive to this choice.  In making this
assumption, we are ignoring the effects of the disk matter which will
retard the WDs as they move to high latitude, an effect that is
relatively small for the high velocity, halo stars of
interest. However, this is a bad approximation for the two disc
components.

\begin{figure*}
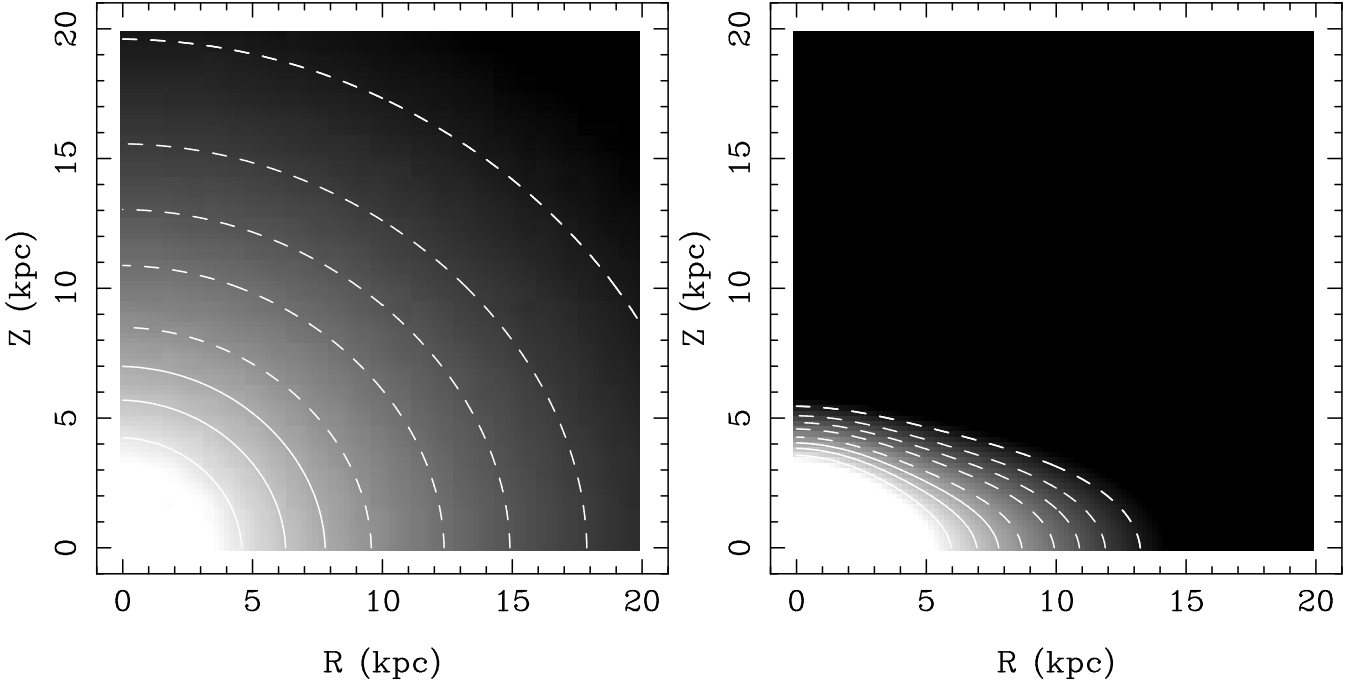

\begin{center}
\leavevmode
\hbox{%
\epsfxsize=0.5\hsize
\epsffile{fig10a.eps}
\hbox{%
\epsfxsize=0.5\hsize
\epsffile{fig10b.eps}}}
\end{center}
\caption{Left: The halo WD density derived from the most likely
BSVDF model (e.g. Table~1) and Eqn.(\ref{eq:densitymodel}). Contours
indicate factors of two increase (solid lines) or decrease (dashed
lines) in density compared to $\rho^{\rm h}_{0,\rm WD}=0.7\times
10^{-4}$~M$_\odot$\,pc$^{-3}$ at the LSR at radius $R$=8~kpc. 
Right: The same for the thick-disc WD density with $\rho^{\rm
td}_{0,\rm WD}=1.1\times 10^{-3}$~M$_\odot$\,pc$^{-3}$.  Note that the true
thick-disc WD density distribution should be more affected by the 
non-spherical disc potential than the halo population, and the 
computed thick-disc density should only be regarded as indicative.}
\end{figure*}

A WD on a high latitude orbit will oscillate in radius between a
minimum and a maximum radius determined by energy and angular momentum
conservation. As the orbit does not close, the ap- and perigalacticons
($r_{{\rm max,min}}$) will eventually become arbitrarily close to the
Galactic disk. Therefore as we are assuming axisymmetry, (and the
orbital precession that would be present in a more realistic
non-spherical potential will obviate the need for this assumption), we
are locally sampling the halo velocity distribution at $(r,\theta)$
for halo velocities $(v_r,v_\theta,v_\phi)$ for which $r_{{\rm
max}}<r_c<r_{{\rm min}}$. For exterior radii, the region of velocity
space that is not sampled is a hyperboloid of revolution of one sheet,
\begin{equation}
(r^2/r_c^2-1)(v_\theta^2+v_\phi^2)-v_r^2\ge2v_c^2\ln(r/r_c).
\end{equation}
For interior radii, the excluded volume is an oblate spheroid
\begin{equation}
v_r^2+(1-r^2/r_c^2)(v_\theta^2+v_\phi^2)\le2v_c^2\ln(r_c/r).
\end{equation}
Two examples of exluded regions are shown in Fig.9.

The connection between the solar neighborhood and the halo is effected
using the constants of the motion (e.g. May \& Binney 1986), for which
a convenient choice is the energy, squared angular momentum and the
angular momentum resolved along the symmetry axis:
\begin{eqnarray}
E  &=&{1\over2}(U^2+(v_c+V)^2+W^2)+\Phi(r_c)\nonumber\\
L^2&=&r_c^2[(v_c+V)^2+W^2]\nonumber\\
L_z&=&r_c(v_c+V).
\end{eqnarray}
This enables us to compute $F(E,L^2,L_z)$ in the halo at
$(r,\theta)$, within the sampled volume of velocity space, using
\begin{eqnarray}
v_r^2&\leftrightarrow&2E-2\Phi(r) -L^2/r^2\nonumber\\
v_\theta^2&\leftrightarrow&L^2/r^2-L_z^2/r^2\sin^2\theta\nonumber\\
v_\phi^2&\leftrightarrow&L_z^2/r^2\sin^2\theta.
\end{eqnarray} 
Making these substitutions, we discover that the velocity distribution
function in the halo at $(r,\theta)$ is also a Gaussian,
\begin{eqnarray}
	F_h(r,\theta;v_r,v_\phi,v_\theta)  = 
	{n_{0,\rm WD}^{\rm h} \over {(2\pi)^{3/2}\sigma_U \sigma_V\sigma_W}} \times 
	\exp\left(-{\Delta\Phi \over \sigma_U^2}\right) \nonumber\\
	\times \exp\left(-v_r^2 \over {2 \sigma_U^2}\right)\times
	\exp\left({-v_\theta^2 \over 2}\left[{{1-x^2} \over \sigma_U^2}+
	{x^2 \over \sigma_W^2}\right]\right) \nonumber \\
	\times \exp\left({-v_\phi^2 \over 2}\left[{{1-x^2} \over {\sigma_U^2}}+
	{{x^2 \sin^2\theta} \over {\sigma_V^2}} + {{x^2 \cos^2\theta} \over
	{\sigma_W^2}}\right]\right) \nonumber \\
	\times \exp\left({{v_\phi(v_c-v_a)x \sin\theta} \over {\sigma_V^2}} -
	{{(v_c-v_a)^2} \over {2\sigma_V^2}}\right),
\end{eqnarray}
where $\Delta\Phi=\Phi(r)-\Phi(r_c)$ is the difference in the
potential between the point $(r,\theta)$ in the halo and the Solar
neighborhood. Our particular choice for the potential gives
$\Delta\Phi=v_c^2\ln(x)$, with $x=r/r_c$. Although this choice for the
potential is particularly convenient, the above VDF for the halo can
be used with any spherical symmetric potential model and locally
constrained Schwarzschild velocity distribution function. Now in order
to compute the density. we must make an assumption about the velocity
distribution function within the excluded volume. The natural
assumption is that it is the same Gaussian function as within the
connected volume. We emphasize that this is probably a pretty good
assumption when the excluded region of velocity space is small, 
though quite suspect
when it is not. Performing the elementary, if tedious, integrals, we
obtain
\begin{eqnarray}\label{eq:densitymodel}
	&n(r,\theta)&={n_h(r_c,\pi/2)\over\sigma_V\sigma_W} x^{-(v_c/\sigma_U)^2}
	\left[{1-x^2\over\sigma_U^2}+
	{x^2\over\sigma_W^2}\right]^{-1/2}\nonumber\\
	&\times&\left[{1-x^2\over\sigma_U^2}+{x^2\sin^2\theta\over\sigma_V^2}
	+{x^2\cos^2\theta\over\sigma_W^2}\right]^{-1/2}\\
	&\times&\exp\left[-
	{(v_a-v_c)^2 \left[{(1-x^2) \over \sigma_U^2}+{x^2\cos^2\theta \over
	\sigma_W^2}\right] \over {{2\sigma_V^2} \left[
	{(1-x^2) \over \sigma_U^2}+{x^2\sin^2\theta \over \sigma_V^2}+
	{x^2\cos^2\theta \over \sigma_W^2}\right]}}\right],\nonumber
\end{eqnarray}
where the factor $(r/r_c)^{-(v_c/\sigma_U)^2}$ comes from the
potential, i.e. $\exp[-{\Delta\Phi/\sigma_U^2}]$=$\exp[-v_c^2/\sigma_U^2 \ln(r/r_c)]$.
This procedure can only be self consistent for $\sigma_U<{\rm Max}[\sigma_V, \sigma_W]$
when $r<\{1-\sigma_U^2/{\rm Max}[\sigma_V^2,\sigma_W^2]\}^{-1/2}$, otherwise the 
computed density is unbounded. (Indeed there appears to be a physical limit on the
vertical velocity dispersion of a halo population within the Galactic plane.)

The density distributions of the halo and thick disc of the most
likely model are plotted in Fig.10. The halo density distribution can
be seen to be roughly in the form of an oblate flattened spheroidal
distribution with axis ratio, $q_{\rm WD}\sim0.9$ at the solar
radius. We expect $q$ is decrease slightly if a more realistic
flattened potential is used, including constributions from the bulge
and disc (\S 7.2). If we assume that the average WD mass is
$0.6$~M$_\odot$, then the mass contained within a shell 0.3--3$\,r_c$
is $\sim$1.5$\times10^9$~M$_\odot$. The total WD mass inside 50~kpc is
$\sim$2.6$\times10^9$~M$_\odot$ for our most likely model.  The radial
variation of density is $n\propto r^{-3.0}$ as argued on the basis of
the fluid model and the mass per octave varies only slowly. This
result shows that the mass fraction of WDs decreases as $\propto
r^{-1.0}$, in the particular potential that we adopted. Of course the
bulge and the disk must certainly be included within $\sim3$~kpc and
we do not expect the circular velocity to be remain roughly constant
beyond $\sim30$~kpc and our model is quite primitive, relative to what
could be created with a more extensive data set and a more
sophisticated form for $\Phi(\vec{r})$.

Even so, if we were to believe the simple potential model, the total
WD mass corresponds to $\sim$0.4\% of the total mass enclosed by
50~kpc, Hence, we find $\Omega_{\rm WD}\sim 10^{-4}$, if $\Omega_{\rm
halo}(<50{\rm ~kpc}) \sim0.03$ (e.g. Bahcall et al. 2000). This
corresponds to $\sim$0.4\% of the total baryon budget in the universe.
Clearly the halo WDs are dynamically unimportant.

\subsection{The Luminosity of Halo White Dwarfs}

A halo mass of $\sim$2.6$\times10^9$~M$_\odot$ in WDs corresponds to
$N_{\rm WD}\sim$4.3$\times10^{9}$ WDs in total. The expected luminosity 
of this population in $R$-band is 
$$ R_{\rm WD}\approx 5 \log({d_{\rm pc}}) + \left<M_R\right> - (5/2)\log N_{\rm WD} - 5 
{\rm ~~~(mag)},$$
where $\left<M_R\right>$ is the average absolute $R$-band magnitude of
the sample of WDs. If we take $\left<M_R\right>\approx 14.3$ from the
sample of observed WDs by Oppenheimer et al. (2001), this reduces to
$R_{\rm WD}\approx 5\log({d_{\rm pc}})-14.8$.  Similarly, if the
luminosity of halo stars is $L_R\sim 4\times 10^7$~L$_\odot$ (BM98;
conveniently assuming Solar colors for the halo stars) and
$M_{R,\odot}$=4.3, one finds $R_{\rm stars}\approx 5\log({d_{\rm
pc}})-20$. If these values are typical for L$_*$ galaxies, then the
integrated luminosity of halo WDs in $R$-band will be $\sim$5 mag
fainter that that of halo stars, both of which are of course
significantly fainter than the stellar disc. We conclude that it will
be exceedingly difficult to observed such a faint halo around external
galaxies. This is made even harder, because this emission is spread
over a much larger solid angle on the sky than for example the stellar
disc.

\subsection{The Origin of Halo White Dwarfs}

We now turn the crucial question: if the density of halo WDs is 
as high as the observations, taken at face value, and our subsequent
calculations indicate, where did all these WDs come from?
The apparent similarity in magnitude, color and age between the
high-velocity halo and low-velocity thick-disc WDs (Fig.1;
Hansen 2001), suggests that the halo WDs originate in the
disc and are subsequently ejected into the halo, through a process
yet unknown to us. Their disc-origin is supported by the low
density of halo stars (e.g. Gould et al. 1998). If halo stars and WDs
come from the same (halo) population, there should be significantly more
stars than WDs in the halo for any reasonable IMF.

One possibility is that binary or triple star 
interactions eject stars into the halo with high velocities (socalled 
runaway stars), although not preferentially WDs. Alternatively, it
is possible that a large population of WDs were born in globular
clusters on radial orbits or satellite galaxies that have been
destroyed by tidal forces over the past 10~Gyr, through which most of
their stars end up in the bulge. Both scenarios are speculative and
also do not satisfactory explain why the ratio of WDs to stars is so
high in the halo.

We propose another possible mechanism, in which most of the
high-velocity WDs orginate from unstable triple stellar
systems. Observations indicate that a high fraction ($\sim$30\%) of
binary stars in the Solar neighborhood have a companion and are in
fact triples (e.g. Petri \& Batten 1965; Batten 1967). If for example
the third outer-orbit star, or one of the binary stars, evolves into a
WD or goes supernova, or if matter can flow between the different
stars (e.g. through Roche-lobe overflows and/or stellar winds) an
initially stable system can change its ratio of orbital periods and at
some point become unstable, if the ratio of orbital periods ($P_{\rm
l}/P_{\rm s}$, the ratio of the largest over smallest orbital period)
exceeds a critical value (e.g. Anosova \& Orlov 1989; Kiseleva,
Eggleton, \& Anosova 1994; Anosova, Colin, \& Kiseleva 1996) that
depends on the mass ratio of the outer star and the inner binary. The
outer-orbit star, often an evolved star itself, can then be ejected
from the triple system with a high spatial velocity (e.g Worrall
1967), giving the inner compact binary system a ``kick'' in opposite
direction, with a typical velocity observed for compact binaries in
the disc (e.g. Iben \& Tutukov 1999) and comparable to the orbital
velocity of the compact binary. A second possibility is that the most
massive star of the inner binary evolves into a WD, without the binary
merging (Iben \& Tutukov 1999), changing the orbital period and making
the system unstable. Also in this case the WD can be ejected,
leaving behind a recoiled binary.

Besides explaining why even young WDs (few Gyr) are observed with high
velocities, a consequence of such a mechanism is that the total mass
in high-velocity WDs (e.g. those in the halo), must be related to the
integrated stellar mass in our Galaxy and the fraction of the stellar
IMF in unstable multiple (e.g. triple) systems. The estimated stellar
mass in our Galaxy is roughly $6\times 10^{10}$~M$_\odot$ (BT87). If
$\sim$15\% of the stellar mass equals the total galactic WD mass,
based on simple population synthesis models (\S 6.1), the inferred
Galactic WD mass should be $\sim9\times 10^9$~M$_\odot$. In \S 6.1, we
estimated from observations that the local WD mass fraction was only
$\sim$5\%, which then corresponds to $\sim3\times 10^9$~M$_\odot$ if
this fraction holds throughout the disc. We estimated $\sim2.6\times 
10^9$~M$_\odot$ for the mass in halo WDs (Sect.7), which then adds to 
a total WD mass of $\sim 6\times 10^9$~M$_\odot$.

There are, however, a number of difficulties associated with the process
described above. First, it implies that one out of every two
WDs must be ejected from the disc, with an average rate of
one WD every $\sim$4 years, for an assumed Galactic age of $\sim10$~Gyr.  
Second, it is not known what the birth rate and
orbital parameters of triple (or higher-order) systems are. In
particular, a large fraction of stellar systems could be intrisically
unstable when they form. Such systems have been observed (e.g. Ressler
\& Barsony 2001), although their ultimate fate is unknown. 
The instability
of triple systems is particularly evident when comparing their orbital
periods (e.g. Tokovinin 1997; Iben \& Tutkov 1999), which shows a
complete absence of triple systems with $P_{\rm l}/P_{\rm s}\la 2.5$
(i.e. the lowest theoretical limit for stability; e.g. Kiseleva 1994),
whereas there is no a priori reason to suspect that systems with these
orbital period ratios are not formed.  A more typical ratio $P_{\rm
l}/P_{\rm s}$ where systems becomes unstable is $\sim$4. Examining the
distribution of $P_{\rm l}/P_{\rm s}$ in more detail reveals some
clustering of triple stellar systems close to the stability limit for
larger orbital periods (Tokovinin
1997; Iben \& Tutukov 1999), although it is not clear whether that could
be a selection effect. For the lower period systems this clustering is 
not the case.  
Third, because evolving triple stars have not been studied in as much
detail as binaries, for example, the velocity of the ejected WDs is
not well known in these processes, although it can be estimated from
analytical models and numerical simulations.

As a rough rule of thumb, stable triple systems can be separated into
a close binary (masses $m_1$ and $m_2$) and a more distant companion
(mass $m_3$) with period ratio exceeding $\sim4$ (see above).  Let us consider two
options. Firstly, suppose that $m_1$ is the most massive star and that
it evolves to fill its Roche lobe and transfer mass, conservatively,
onto $m_2$. This will initially shrink the orbit and increase the
period ratio until $m_1=m_2$ when the separation is $a_{{\rm min}}$.
Further mass transfer will then lead to an increase of the close
binary orbit and a reduction of the period ratio $\propto
m_1^{-3}m_2^{-3}$.  As the close binary orbit expands, the proto--WD
speed $v_1$ will initially increase $\propto m_1m_2^2$ and then
decrease. For example, with $m_1=2$~M$_\odot$, $m_2=1$~M$_\odot$
initially the close binary period could decrease by a factor $\sim3$
by the time $m_1\sim0.6$~M$_\odot$ and, provided that the companion's
orbital radius $\la 6\, a_{{\rm min}}$, the orbit can become unstable
at this point. If $m_1$ is ejected by a slingshot effect involving
$m_3$, then a characteristic velocity at infinity is $v_1\sim200\cdot
(a_{{\rm min}}/4{\rm R}_\odot)^{-1/2}$.  Higher speeds are possible
with favorable orbits, especially retrograde companion
orbits. Secondly, suppose that $m_3>m_1+m_2$. The mass transfer is
less likely to be conservative although there is likely to be a net
loss of orbital angular momentum associated with the loss of mass.
The net result is that the outer binary orbit will initially shrink,
again promoting instability.

General analytical and numerical analyses of triple-star systems,
randomly placed in a spherical volume, show that most ($\ga$80\%)
systems are unstable and typically eject the lowest mass star with a
median velocity comparable to their orbital velocity (e.g. Standish
1972; Monaghan 1976a,b; Sterzik \& Durisen 1998), as also indicated
above. A more detailed analysis by Sterzik \& Durisen (1998), where
the ejection velocity distribution is weighted by the stellar mass
function, shows that $\sim$10 percent of the ejected stars (e.g. WDs)
have velocities $\ga$3 times the typical orbital velocity. Given their
stellar mass function, the typical mass of an evolved triple-star
systems is then $\sim$5~M$_\odot$. As we saw previously, if the
orbital radius is of the order of the stellar radius (i.e. orbital
periods of several days), escape velocities of several hundred \kms\
can be reached. For orbital periods of several hundred days
(e.g. Tokovinin 1997), the typical orbital velocity will still be
$\sim$80~\kms. WDs have to gain a velocity of $\sim$100--150~\kms\ (in
addition to their orbital velocity and velocity with respect to their
LSR) to atain velocities comparable to the observed high-velocity
WDs. Hence, the proposed mechanism quite easily reaches the required
velocities. Even in those systems with longer orbital periods
($\sim$1000 days), several tens of percent of the escaping stars will
atain very high velocities (e.g. Sterzik \& Durisen~1998).

Additional support for the notion that these processes might be
quite common, could be the good agreement which Valtonen (1998) recently
found between the observed mass ratio of binaries and that determined
from a statistical theory of three-body disruption (Monaghan 1976a,b).
This might indicate that many stars are born in triple or higher-order 
systems, which become unstable and eject stars until only a stable binary 
is left.

In order to explore this hypothesis further, we have constructed a 
relatively simple model
of the Galaxy as follows; we suppose that the mass comprises a spherical
halo
\begin{equation}
	\rho_{\rm h}(r)={(190~{\rm km~s^{-1}})^2 \over {4 \pi G}}
	{{3 r_{\rm h}^2 + r^2} \over {(r_{\rm h}^2 + r^2)^2}},
\end{equation}
with $r_{\rm h}=8$~kpc, which generates a logarithmic potential, and a
disc
\begin{equation}
	\rho_{\rm d}(R)= 60\cdot e^{-R/h_{\rm d}}\delta(z)\mbox{~~M}_\odot\,{\rm pc}^{-3},
\end{equation}
with $h_{\rm d}=3$~kpc and a bulge, which we for simplicity flatten onto the 
Galactic plane,
\begin{equation}
	\rho_{\rm b}(R)= 190\cdot e^{-R/h_{\rm b}}\delta(z)\mbox{~~M}_\odot\,{\rm pc}^{-3},
\end{equation}
with $h_{\rm b}=0.8$~kpc. The combination of these potentials gives a nearly 
flat rotation curve beyond several kpc (e.g. Dehnen \& Binney 1998a), 
with circular velocity $v_c\approx 220$~\kms. As the potential is
no longer spherically symmetric, the total angular momentum is not an integral of
motion. 

Given this model of the Galaxy, it is then a straighforward matter to
follow the trajectories of observed WDs back in. In order to measure
the contemporary WD source velocity distribution (per unit disk mass,
per unit time, per unit speed), which we denote by $g(v)$, we must
make some assumptions.  To keep matters simple, we suppose that $g(v)$
is independent of radius and decreases with lookback time,
$t\propto\exp[-t/\tau]$, where $\tau=7.5{\rm Gyr}$ as indicated by
the data exhibited in Fig.~1. As individual orbits quickly sample a
large range of disk radii it is not easy to derive the radial launch
rate from the local WD distribution function. The best approach
utilizing the present data set is to fit the halo distribution function 
from Table.~1.

We use again Jeans' theorem to relate the observed VDF $F^h(\vec
x_0,\vec v_0)$ to that at the source. Adopt one point in local
velocity space and relate to $g(v)$ through
\begin{equation}
F_i\equiv F^h(\vec x_0, \vec v_{0i})=\sum_{j=1}^m W_{ij}\,g[v_j(v_{0i})],
\end{equation}
where the sum takes account of $m$ disk crossings along the orbit.
The weights $W_{ij}$ are given by
\begin{equation}
W_{ij}=\frac{\Sigma(r_j)\, \exp[-{t_j/\tau}]}{4\pi v_j^2v_{jz}}
\end{equation}
and $r_j$, $v_j$ and $t_j$ are evaluated at the j-th disk crossing.
We next discretize $g(v_j)$ using 50~\kms\ bins and develop a least
squares solution to Eq.(22)
\begin{equation}
g_j=\frac{\sum_{i,j} W_{ij}F_i}{\sum_{i=1}^N W_{ik}W_{kj}}
\end{equation}
where we sample the distribution function at N random points.

The result indicates that the mean speed of the WD at launch is very
high, $\bar v$=250--300~km s$^{-1}$, with a birth rate of
$\sim$10$^{-4}$~(M$_\odot$~\kms~Gyr)$^{-1}$ at these velocities,
consistent with $\sim$10\% of the stellar mass in the Galactic disc
having been transformed in WDs. The low rotational velocity of the
observed WDs requires them to originate in the inner 2--3~kpc of the
Galactic disk, so they loose most of their rotational velocity when
climbing out of the Galactic potential to the Solar neighborhood.
The latter result is not unexpected, because most  stellar mass 
($\sim$75\% for a disk scale length of $h\sim$3~kpc) lies inside the 
solar radius and in addition ${\rm d}M/{\rm d}r\propto r\exp(-r/h)$ peaks 
at $r=h$. Hence a typical WD is expected to originate 
in the disc at $r=h\sim3$~kpc.

These calculations are consistent with, though do not require, that
the WDs are a product of stellar evolution in the disk. Three lines of
evidence support this conclusion. (i) The age distribution is just
about what would be expected from a population of main sequence stars
with a reasonable mass function from $\sim$0.8~M$_\odot$ to
$\sim$6~M$_\odot$. The stars are born at a rate consistent with
stellar evolution. The most populous stars are of mass
$\sim1-2$~M$_\odot$ and take $\sim5-10$~Gyr to evolve. Even if the
Galaxy is $\sim14$~Gyr old, we do not expect a large proportion of
stars with ages $\ga10$~Gyr. A more careful, Galactic synthesis model
is consistent with these statements.  (ii) The velocities are
isotropic with no discernible asymmetric drift.  This is just what
would be expected from a disk population that launches WDs in random
directions.  The median speed observed is $\sim$200--300~km s$^{-1}$. (iii)
The radial variation of white-dwarf sources appears to fall with
radius in a similar fashion to the the stellar distribution.
 
Of course, the situation is far more complex than this. However, the
similarities between the characteristic escape speed and the halo
velocity dispersion, together with the shortage of viable alternatives
makes this a scenario that is worth investigating further. But it 
should again be emphasized that a much larger sample of halo WDs is 
needed to place the above simple estimates on a more firm basis.

\section{Microlensing Optical Depths}

The next question to ask is whether the population of WDs
in the halo can explain the observed microlensing events towards the Large
Magellanic Cloud, with optical depth $\tau_{\rm LMC}=1.2\times
10^{-7}$ (Alcock et al. 2000). Using the local density and the global
density profile of the population of WDs in the halo, derived
in the previous sections, it is possible to integrate over the density
profile along the line-of-sight toward to LMC ($\ell=280.5^\circ$,
$b=-32.9^\circ$ and $D=50$~kpc) and obtain the optical depth for any
arbitrary set of parameters for the local VDF.

For the most likely model, listed in Table~1 (full data set), we
compute a microlensing optical depth of the halo WDs, $\tau^{\rm
h}_{\rm WD}\approx1.3\times 10^{-9}$. The optical depth of the
thick disc is $\tau^{\rm td}_{\rm WD} \approx4\times 10^{-9}$.  Hence,
even in the most optimistic case, the integrated thick-disc plus halo
WDs have an optical depth that is still 1--2 orders of magnitude below
that observed. This difference can not be made up by a, most likely
small, incompleteness of high-velocity WDs.

If we scale the results in Section 6 and 7 to the LMC itself, assuming
(i) a similar shape of the velocity ellipsoid, (ii) velocities that
scale with a factor $(v_c^{\rm LMC}/v_c^{\rm Gal})=(72/220)$
($v_c^{\rm LMC}=72$~\kms; e.g. Alves \& Nelson 2000) and distances
that scale as $v_c^2/r$=constant, and (iii) the ratio of total WD mass
to stellar mass in the LMC is similar to that inferred in our Galaxy
($\sim$9\% from $M_{\rm stars}^{\rm LMC}=5.3\times 10^9$~M$_\odot$;
e.g. Alves \& Nelson 2000), we find a self-lensing microlensing
optical depth for the center of the LMC of $\tau\sim 2\times 10^{-8}$,
lower than the results from the MACHO collaboration (Alcock et
al. 2000). We find that the results are quite sensitive to the
assumped parameters, especially the velocity dispersion. The WD
density is $\propto r^{-(v_c/\sigma_U)^2}$ and gets unbounded if the
velocity dispersion is lower somewhat. This illustrates that a
``shroud'' of white dwarfs around the LMC (e.g. Evans \& Kerins 2000),
in principle, could account for the observed optical depth, although
it does require that 50\% of the halo WD mass is concentrated in the
inner 0.7~kpc and rather fine-tuned model parameters. The latter could
be overcome if the mass profile of the LMC has a finite core.

A similarly rough scaling for the edge-on spiral gravitational lens
B1600+434 -- where we assume 4 times more mass in WDs than in
B1600+434 -- suggests $\tau_{\rm WD}\sim10^{-3}$ along the
line-of-sight towards lensed image A, at 6~kpc above the galaxy
plane. This optical depth appears insufficient to explain the apparent
radio-microlensing events in this system (e.g. Koopmans \& de Bruyn
2000; Koopmans et al.  2001). We stress the large uncertainties
associated with these extrapolated calculations.

\begin{figure}
\begin{center}
\leavevmode
\hbox{%
\epsfxsize=\hsize
\epsffile{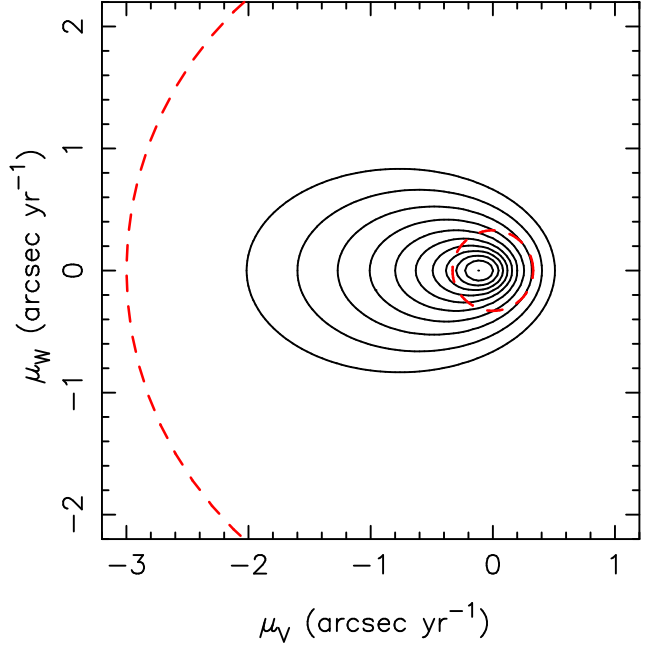}}
\end{center}
\caption{The probability density of thick-disc plus halo WD proper
motions towards the Galactic anti-center (see text). The contours are
spaced by factors of two, decreasing from the maximum. The inner
dashed circle indicates the proper-motion limit of 0.33~\muy, whereas the
outer dashed circle is indicative of 3.0~\muy. The extension towards
negative $\mu_V$ exemplifies the asymmetric drift. The oblateness of
the probability contours is related to the ratio $\sigma_W/\sigma_U$.}
\end{figure}

\section{Future Surveys}

As we have discussed, the main reason why it is hard to determine the
halo density of the observed WD is that the sample surrounds 
the SGP and that we have poor resolution of the vertical velocity $W$. 
In order to choose between various possible local distribution functions, 
it is necessary to repeat the survey at lower Galactic latitude. We
discuss two potential surveys for the near future that allow a rigorous
test of some of these underlying assumptions.

\subsection{The Galactic Anti-Center}

In Fig.11, we show the probability distribution of proper motions, 
seen in the direction of the Galactic anti-center ($\ell=180^\circ$ and
$b=0^\circ$). For simplicity we adopt the same survey limits as
used by Oppenheimer et al (2001) and a standard-candle WD with
an average absolute magnitude $M_{R59F}=14.3$. The latter translates into
a maximum survey depth of $\sim$125~pc. We used our most-likely model
in the calculation. If a WD survey in this particular direction shows a 
proper-motion distribution similar to that in Fig.11, it would be strong
support to the notion that the WDs with high velocities in the $U-V$ 
plane also have high velocities in the $W$--direction and therefore form 
a genuine halo population. A significantly flattened proper-motion 
distribution would support a thick-disc interpretation of the 
sample found by Oppenheimer et al. (2001), but requires a rather
artificial VDF.

\subsection{Halo White Dwarfs with the ACS}

A new opportunity to measure WD proper motions is presented by the
Advanced Camera for Surveys\footnote{http://www.stsci.edu/cgi-bin/acs} 
(ACS). It is likely that it will be
proper-motion limited, rather than magnitude limited. Scaling from the
astrometry carried out by Ibata et al. (1999) on the HDF, and
requiring that the transverse velocities be measured to an accuracy
$\sim$30~\kms, we estimate that a ``wide'' survey to a depth 
$\sim$1.2~mag shallower than the HDF(N) could be carried out to
a limit 
$$ \mu^{\rm wide}_{\rm lim}\sim 3.0\,d_{\rm kpc}^2 
\times 10^{0.4(M_R-15)}{\rm ~mas~yr}^{-1} $$
over $\Omega_{\rm S}\sim10^{-4}$~ster rad ($\sim$0.3 square degree) and a ``deep'' 
survey could reach 
$$ \mu^{\rm deep}_{\rm lim}\sim 0.3\,d_{\rm kpc}^2 \times
10^{0.4(M_R-15)}{\rm ~mas~yr}^{-1},$$ to a depth $\sim1.2$~mag deeper
than the HDF(N) over $\Omega_{\rm S}\sim10^{-6}$~ster rad
($\sim$120 square arcmin). The expected number of halo WDs in these fields is
\begin{equation}
	N^{\rm halo}_{\rm WD} = \Omega_{\rm S}
	\int {\rm d}\vec{v}^3 \int_0^{v_\perp/\mu_{\rm lim}} {\rm d}s\, s^2 
	F_h(r,\theta;v_r,v_\phi,v_\theta),
\end{equation}
where $s$ measures the distance along the line of sight.  Evaluating
this integral for the direction of the HDF(N), we find that $N^{\rm halo}_{\rm
WD}\approx10\cdot \left< 10^{0.3(15-M_{R})}\right>$ for the wide
survey and $N^{\rm halo}_{\rm WD}\approx0.6\cdot \left<
10^{0.3(15-M_{R})}\right>$ for the deep survey. Evaluating the
luminosity for the high-velocity WDs, we find that the number density
per magnitude is roughly constant in the range
13$<$$M_R$$<$16~mag. This gives us
$$ N^{\rm halo}_{\rm WD}\approx 5 {\rm ~~~(wide)~~~or~~~} 
N^{\rm halo}_{\rm WD}\approx 0.3 {\rm ~~~(deep)}. $$ The wide survey will
typically observe to a distance of $\sim$0.5~kpc, whereas the deep
survey should see to $\sim$1~kpc. The expected number in the existing
WFPC2 observations of the HDF(N) is $\sim$0.1 and the negative report
of Richer (2001) is not surprising.

Of course, these calculations are only illustrative. More WDs are
expected to be found associated with the thick disc and would be seen
along lines of sight that look closer to the inner Galaxy. However, it
is clear that if we can measure the white-dwarf colors and proper
motions over a few directions, then we can use the distribution
function approach, outlined above, to measure the WD density
throughout a major portion of the halo.

\section{Discussion}

We have analysed the results of Oppenheimer et al. (2001) and argued
that there appear to be at least two kinematically distinct
populations of WDs (at the $>$99\% C.L.), which one might term a thick
disc and a flattened halo population, respectively. In contrast, these
populations are indistinguishable in their luminosity, color and age
distribution. Our most likely model indicates that the halo WDs
constitute $\sim$0.8~percent of the nominal local halo density,
although this fraction decreases as $\rho^{\rm h}_{\rm WD}/\rho^{\rm
h}_{\rm total}$$\approx 0.008\cdot (r/r_c)^{-1.0}$ with increasing
Galacto-centric radius $r/r_c$, for an assumed Galactic density
profile $\rho_{\rm h}\propto r^{-2}$.  A more conservative lower limit
of 0.3 percent (90\% C.L.) can be placed on the local halo mass
fraction in WDs. This WD density is 5.0$^{+9.5}_{-3.2}$ (90\% C.L.)
times higher than previously thought and comparable to the local
stellar halo density (Gould et al. 1998). This is in conflict with the
estimates of WD densities, based on population synthesis models and
Salpeter initial mass functions, and requires a non-standard
explanation if one requires these WDs to be formed in the halo. Based
on the notion that 90\% of the WDs, found by Oppenheimer et
al. (2001), are proper-motion limited (Sect.6), we do not expect the
white-dwarf luminosity function to rise sharply beyond the survey
magnitude limit (i.e. WDs with ages $>$10~Gyr), although a distinct
very faint population of WDs with ages $\gg$10~Gyr, which are not
related to the presently observed population, can not be excluded.

We find that the halo WD population is flattened
($q=(c/a)_\rho\sim0.9$), and expect $q$ to become somewhat smaller
once a more realistic flattened potential of the disk, bulge and halo
is used in the models. The value for $q$ that we quote should be
regarded as an upper limit.

The microlensing optical depth inferred from the halo WD population is
estimated at $\tau^{\rm h}_{\rm WD}\approx1.3\times10^{-9}$. Even if
we include the contribution from the thick disk, the optical depth
remains $\la$5\% of that observed toward the LMC ($\tau_{\rm
LMC}\approx1.2\times10^{-7}$; Alcock et al. 2000).  We conclude that
the population of halo WDs, observed by Oppenheimer et al. (2001) and
on orbits that are probed by the velocity distribution function in the
Solar neighborhood, do not contribute significantly to the observed
microlensing events toward the LMC.  One avoids this, if either
$\sim$25 times more WDs are on orbits with perigalacticons well
outside the Solar neighborhood (for example a mass shell) or on
resonant orbits. Both types of orbits could be undersampled
locally. Similarly, as mentioned above, if there exists yet another
population of very faint old ($\sim$15~Gyr) WDs in the halo, they
could also increase the number of microlensing events towards the
LMC. However, we regard both solutions as unlikely and artificial.  In
addition, both solutions would violate constraints from metal
abundances (see Sect.1) and (simple) population synthesis models.  In
Sect.8, we showed that a similar ``shroud'' of white dwarfs around the
LMC itself, in principle, can account for the observed optical depth,
although the result is strongly model dependent.

We have performed a number of consistency checks of the data. For
example, we find excellent agreement between the maximum-likelihood
value of the asymmetric drift, the statistical value and the value
inferred from other stellar population in the solar neighborhood.
There is also no evidence for either a drift of the WD population in
the $U$ or $W$ direction, which could have indicated that the
population was not in dynamic equilibrium or could be part of a
stellar stream or group. In both cases our approach, based on the
collisionless Boltzmann (or Vlasov) equation, would have been invalid.
In addition, the thick-disc likelihood results agree well with the
empirical relation between the asymmetric drift ($v_a$) and the radial
velocity dispersion ($\sigma_U$), determined from a range of local
stellar populations. We also find that the asymmetric drift of the
halo population is close to that (i.e. $v_a \sim v_c$) expected for a
population that is mostly pressure supported (i.e. $\sigma_U\sim
v_c/\sqrt{2}$). If we assume that $\sigma_U$=$\sigma_V$ for the WD
population, the likelihood decreases by a factor $\sim10^6$ and the
thick-disc result no longer agree with the empiral relation. A
flattened velocity ellipsoid ($\sigma_U$$<$$\sigma_V$) naturally
explains the relative scarcity of WDs in the regions around
$(U,V)=(0,-220)$~\kms\ and $V<-220$~\kms, noticed by Reid et
al. (2001).

The absence of any clear correlations of velocity with color, absolute
magnitude or age, and the dynamical self-consistency checks give us
confidence that the population of WDs found by Oppenheimer et
al. (2001) does not contain a significantly hidden bias. Especially,
the fact that the survey is 90\% proper-motion limited, not
magnitude-limited, underlines that we do not miss many WDs that are
too faint (but see the comments above). The range of luminosities
should thus provide a reasonable representation of the underlying
luminosity function of this particular population of halo WDs.

These internal and external consistency checks should make our results
quite robust. However, there are uncertainties associated with errors
in the photometric parallaxes which feed into both the densities and
velocities. In addition, several of the WDs in this sample might
belong to binaries, although an inspection of the two-color diagram
(Fig.1) suggests that also this is unlikely to be a major
problem. Only a systematic overestimate of distances by a factor
$\ga$2.5 would move most high-velocity WDs into the thick-disc regime,
but would also result in a typical velocity dispersion of the sample
only half that of the thick disc and average WD luminosities that are
too small. The absence of apparent correlations with the WD
velocities, makes it unlikely that distance overestimates would
preferentially occur for the highest velocity WDs. Another way out of
the conclusion that there exists a significant population of halo WDs,
would be to accept that the velocity ellipsoid of these high
proper-motion WDs is significantly flattened in the vertical
direction.  Although, this would be consistent with the observations,
as we indicated in Sect.5, it would require a highly unusual VDF,
unlike what is observed for both young and old stellar populations in
the Solar neighborhood.

We have also provided further evidence that the fast and slow WDs come
from the same population. This would not be a surprise if they were
mostly born in the disk and then deflected dynamically to high
altitude.  We have proposed a possible mechanism that could
preferentially eject WDs from the disc, involving orbital
instabilities in evolving multiple (e.g. triple)  stellar systems (\S
7.2). We showed that the total halo plus disc WD mass of the Galaxy
(i.e. $\sim$15\% of its stellar mass) is roughly consistent with that
expected from the total stellar mass in our Galaxy, based on standard
population synthesis models. The agreement suggests that not a
re-thinking of Galactic starformation models is required, but that the
real question is how to eject disc WDs into the halo with high
velocities. We propose that this can be achieved through the orbital
instabilities in evolving multiple stellar systems.

The key to understanding the WD distribution with more confidence is
undoubtedly to measure the component of velocity perpendicular to the
Galactic disk and this is best accomplished with a proper motion
survey at lower latitude. We have suggested two possible surveys for the near
future: one towards to Galactic anti-center, similar to the survey
discussed in this paper, and one that could be done with the Advanced
Camera for Surveys (ACS) on HST. The first survey could unambiguously
show that the velocity dispersion perpendicular to the disc is
comparable to that parallel to the disc.  Measuring the density of WDs as a
function of Galacto-centric radius (for example with the ACS), together
with a local WD density calibration, can provide strong constraints on
the shape of the Galactic potential, as is apparent from Eqns (17) and
(18) in Sect.7.  These constraints improve dramatically if we can also constrain
the VDF at the same points in the halo. This is especially relevant in
case of the halo WDs, even though they are faint, because their high
velocity dispersion allows one to probe large (tens of kpc) Galacto-centric
distances and still be in dynamic contact with the Solar
neighborhood. The relation between the VDFs at different radii is then
given through the integrals of motion that are conserved along orbits
between these points and the Galactic potential. Measuring the
density and VDFs along these orbits can then, in principle, be used to 
reconstruct the Galactic potential. It is also surprising that the scale height of
WDs, kinematically belonging to the thin disc, appears at least
twice that previously thought (Majewski \& Siegel 2001). Could 
this population form a bridge to a flattened white-dwarf halo population?

We conclude that the discovery of a surprisingly large population of
high velocity, old WDs, made possible by advances in understanding
their atmospheres and colors, is a significant accomplishment and
seems to opens up a new unexpected window into the stellar archaeology
of our Galaxy. However, to confirm or reject the ideas put forward 
by Oppenheimer et al. (2001) and in this paper a significantly large
sample is required, not only out of the Galactic plance, but also
in the Galactic plane. In addition deeper surveys (e.g. with the ACS)
could probe much farther into the halo and possibly even detect WDs that
are significantly fainter. Deeper surveys would also be able to
probe the transition between thick-disc and halo and be less 
`contaminted' by disc WDs.

%%%%%%%%%%%%%%%%%%%%%%%%%%%%%%%%%%%%%%%%%%%%%%%%%%%%%%%%%%%%%%%%%%%%%%%%%%%%%%%%

\section*{Acknowledgments}

The authors thank Ben Oppenheimer for many valuable discussions and
providing tables with their results prior to publication. LVEK thanks
Dave Chernoff for several discussions. The authors are indebted to
David Graff and Andy Gould for bringing our attention to a mistake in
the normalisation of the likelihood function. This research has been
supported by NSF~AST--9900866 and STScI~GO--06543.03--95A.

{}

\end{document}